\pdfoutput=1

\documentclass[12pt,a4paper,final]{article}
 
\usepackage{ifthen} 
\usepackage{subcaption}
\usepackage{graphicx}
\usepackage{tabulary}
\usepackage{tabularx}
\usepackage{amsmath}
\newboolean{pdflatex}
\setboolean{pdflatex}{true} 

\usepackage{float}
\floatstyle{plaintop}
\restylefloat{table}
\usepackage[tableposition=top,font={small},labelfont={normalfont,small}]{caption}
\usepackage{placeins}
\usepackage{pdflscape}

\usepackage{afterpage}
\newboolean{articletitles}
\setboolean{articletitles}{true} 

\newboolean{uprightparticles}
\setboolean{uprightparticles}{false} 

\renewcommand{\textfraction}{0.01}
\renewcommand{\floatpagefraction}{0.99}
\renewcommand{\topfraction}{0.9}
\renewcommand{\bottomfraction}{0.9}

\usepackage{microtype}
\usepackage{lineno}  
\usepackage{xspace} 

\usepackage{graphicx}  
\usepackage{color}
\usepackage{colortbl}
\graphicspath{{././}} 

\usepackage{amsmath} 
\usepackage{amssymb}
\usepackage{amsfonts}
\usepackage{upgreek} 

\newcommand*\patchAmsMathEnvironmentForLineno[1]{%
\expandafter\let\csname old#1\expandafter\endcsname\csname #1\endcsname
\expandafter\let\csname oldend#1\expandafter\endcsname\csname
end#1\endcsname
 \renewenvironment{#1}%
   {\linenomath\csname old#1\endcsname}%
   {\csname oldend#1\endcsname\endlinenomath}%
}
\newcommand*\patchBothAmsMathEnvironmentsForLineno[1]{%
  \patchAmsMathEnvironmentForLineno{#1}%
  \patchAmsMathEnvironmentForLineno{#1*}%
}
\AtBeginDocument{%
\patchBothAmsMathEnvironmentsForLineno{equation}%
\patchBothAmsMathEnvironmentsForLineno{align}%
\patchBothAmsMathEnvironmentsForLineno{flalign}%
\patchBothAmsMathEnvironmentsForLineno{alignat}%
\patchBothAmsMathEnvironmentsForLineno{gather}%
\patchBothAmsMathEnvironmentsForLineno{multline}%
}

\usepackage{hyperref}    
\usepackage[all]{hypcap} 

\usepackage{xspace} 
\usepackage{upgreek}

\def\lhcb {\mbox{LHCb}\xspace}

\def\MagUp {\mbox{\em Mag\kern -0.05em Up}\xspace}

\ifthenelse{\boolean{uprightparticles}}%
{

 \def\Pepsilon    {\ensuremath{\upepsilon}\xspace}

 \def\Ppi         {\ensuremath{\uppi}\xspace}

 \def\Ppsi        {\ensuremath{\uppsi}\xspace}

 \def\PDelta      {\ensuremath{\Delta}\xspace}                 
 \def\PXi      {\ensuremath{\Xi}\xspace}                 
 \def\PLambda      {\ensuremath{\Lambda}\xspace}                 
 \def\PSigma      {\ensuremath{\Sigma}\xspace}                 
 \def\POmega      {\ensuremath{\Omega}\xspace}                 
 \def\PUpsilon      {\ensuremath{\Upsilon}\xspace}                 
 
 \def\PB      {\ensuremath{\mathrm{B}}\xspace}                 
                  
 \def\PD      {\ensuremath{\mathrm{D}}\xspace}

 \def\PJ      {\ensuremath{\mathrm{J}}\xspace}                 
 \def\PK      {\ensuremath{\mathrm{K}}\xspace}

 \def\PN      {\ensuremath{\mathrm{N}}\xspace}                 
                  
 \def\PP      {\ensuremath{\mathrm{P}}\xspace}

 \def\PX      {\ensuremath{\mathrm{X}}\xspace}                 
                  
 \def\PZ      {\ensuremath{\mathrm{Z}}\xspace}                 
                  
 \def\Pb      {\ensuremath{\mathrm{b}}\xspace}                 
 \def\Pc      {\ensuremath{\mathrm{c}}\xspace}

 \def\Pi      {\ensuremath{\mathrm{i}}\xspace}

 \def\Pm      {\ensuremath{\mathrm{m}}\xspace}

}
{

 \def\Pepsilon    {\ensuremath{\epsilon}\xspace}

 \def\Ppi         {\ensuremath{\pi}\xspace}

 \def\Ppsi        {\ensuremath{\psi}\xspace}                 
                  
 \mathchardef\PDelta="7101
 \mathchardef\PXi="7104
 \mathchardef\PLambda="7103
 \mathchardef\PSigma="7106
 \mathchardef\POmega="710A
 \mathchardef\PUpsilon="7107
                  
 \def\PB      {\ensuremath{B}\xspace}                 
                  
 \def\PD      {\ensuremath{D}\xspace}

 \def\PJ      {\ensuremath{J}\xspace}                 
 \def\PK      {\ensuremath{K}\xspace}

 \def\PN      {\ensuremath{N}\xspace}                 
                  
 \def\PP      {\ensuremath{P}\xspace}

 \def\PX      {\ensuremath{X}\xspace}                 
                  
 \def\PZ      {\ensuremath{Z}\xspace}                 
                  
 \def\Pb      {\ensuremath{b}\xspace}                 
 \def\Pc      {\ensuremath{c}\xspace}

 \def\Pi      {\ensuremath{i}\xspace}

 \def\Pm      {\ensuremath{m}\xspace}

}

\makeatletter
\ifcase \@ptsize \relax
  \newcommand{\miniscule}{\@setfontsize\miniscule{4}{5}}
\or
  \newcommand{\miniscule}{\@setfontsize\miniscule{5}{6}}
\or
  \newcommand{\miniscule}{\@setfontsize\miniscule{5}{6}}
\fi
\makeatother

\DeclareRobustCommand{\optbar}[1]{\shortstack{{\miniscule (\rule[.5ex]{1.25em}{.18mm})}
  \\ [-.7ex] $#1$}}

\def\cquark    {{\ensuremath{\Pc}}\xspace}

\def\bquark    {{\ensuremath{\Pb}}\xspace}

\def\pion   {{\ensuremath{\Ppi}}\xspace}

\def\pip    {{\ensuremath{\pion^+}}\xspace}
\def\pim    {{\ensuremath{\pion^-}}\xspace}

\def\kaon    {{\ensuremath{\PK}}\xspace}
  \def\Kbar    {{\kern 0.2em\overline{\kern -0.2em \PK}{}}\xspace}

\def\KorKbar    {\kern 0.18em\optbar{\kern -0.18em K}{}\xspace}

\def\Kp      {{\ensuremath{\kaon^+}}\xspace}
\def\Km      {{\ensuremath{\kaon^-}}\xspace}

  \def\Dbar    {{\kern 0.2em\overline{\kern -0.2em \PD}{}}\xspace}

\def\DorDbar    {\kern 0.18em\optbar{\kern -0.18em D}{}\xspace}

\def\B       {{\ensuremath{\PB}}\xspace}
\def\Bbar    {{\ensuremath{\kern 0.18em\overline{\kern -0.18em \PB}{}}}\xspace}

\def\BorBbar    {\kern 0.18em\optbar{\kern -0.18em B}{}\xspace}
\def\Bz      {{\ensuremath{\B^0}}\xspace}

\def\Bu      {{\ensuremath{\B^+}}\xspace}

\def\Bp      {{\ensuremath{\Bu}}\xspace}

\def\jpsi     {{\ensuremath{{\PJ\mskip -3mu/\mskip -2mu\Ppsi\mskip 2mu}}}\xspace}
\def\psitwos  {{\ensuremath{\Ppsi{(2S)}}}\xspace}

  \def\Y#1S{\ensuremath{\PUpsilon{(#1S)}}\xspace}

\def\Lbar        {{\ensuremath{\kern 0.1em\overline{\kern -0.1em\PLambda}}}\xspace}
\def\LorLbar    {\kern 0.18em\optbar{\kern -0.18em \PLambda}{}\xspace}

\def\BR         {\BF}

\def\to                 {\ensuremath{\rightarrow}\xspace}

\def\AT#1     {\ensuremath{A_{\mathrm{T}}^{#1}}\xspace}           

\def\C#1      {\ensuremath{\mathcal{C}_{#1}}\xspace}                       
\def\Cp#1     {\ensuremath{\mathcal{C}_{#1}^{'}}\xspace}                    
\def\Ceff#1   {\ensuremath{\mathcal{C}_{#1}^{\mathrm{(eff)}}}\xspace}        
\def\Cpeff#1  {\ensuremath{\mathcal{C}_{#1}^{'\mathrm{(eff)}}}\xspace}       
\def\Ope#1    {\ensuremath{\mathcal{O}_{#1}}\xspace}                       
\def\Opep#1   {\ensuremath{\mathcal{O}_{#1}^{'}}\xspace}                    

\newcommand{\tev}{\ensuremath{\mathrm{\,Te\kern -0.1em V}}\xspace}
\newcommand{\gev}{\ensuremath{\mathrm{\,Ge\kern -0.1em V}}\xspace}
\newcommand{\mev}{\ensuremath{\mathrm{\,Me\kern -0.1em V}}\xspace}
\newcommand{\kev}{\ensuremath{\mathrm{\,ke\kern -0.1em V}}\xspace}
\newcommand{\ev}{\ensuremath{\mathrm{\,e\kern -0.1em V}}\xspace}
\newcommand{\gevc}{\ensuremath{{\mathrm{\,Ge\kern -0.1em V\!/}c}}\xspace}
\newcommand{\mevc}{\ensuremath{{\mathrm{\,Me\kern -0.1em V\!/}c}}\xspace}
\newcommand{\gevcc}{\ensuremath{{\mathrm{\,Ge\kern -0.1em V\!/}c^2}}\xspace}
\newcommand{\gevgevcccc}{\ensuremath{{\mathrm{\,Ge\kern -0.1em V^2\!/}c^4}}\xspace}
\newcommand{\mevcc}{\ensuremath{{\mathrm{\,Me\kern -0.1em V\!/}c^2}}\xspace}

\def\mum  {\ensuremath{{\,\upmu\mathrm{m}}}\xspace}

\def\invfb   {\ensuremath{\mbox{\,fb}^{-1}}\xspace}

\def\ps   {\ensuremath{{\mathrm{ \,ps}}}\xspace}

\def\gsim{{~\raise.15em\hbox{$>$}\kern-.85em
          \lower.35em\hbox{$\sim$}~}\xspace}
\def\lsim{{~\raise.15em\hbox{$<$}\kern-.85em
          \lower.35em\hbox{$\sim$}~}\xspace}

\def\PDF {PDF\xspace}

\def\ptot       {\mbox{$p$}\xspace}
\def\pt         {\mbox{$p_{\mathrm{ T}}$}\xspace}

\def\degrees{\ensuremath{^{\circ}}\xspace}

\def\tell1  {TELL1\xspace}
\def\ukl1   {UKL1\xspace}

\usepackage{cite} 
\usepackage{mciteplus}

\usepackage{flafter}
\usepackage{siunitx}
\usepackage{datetime}

\hypersetup{pdfinfo={
Author={The LHCb collaboration },
Title={A model-independent confirmation of the Z(4430)- state},
Keywords={Z(4430)-; LHCb; Charmonium-like; Exotic states},
CreationDate={20151007},
Subject={Abstract: The decay B0 -> psi(2S)K-pi- is analyzed using 3/fb of pp collision data collected
with the LHCb detector. A model-independent description of the psi(2S)pi mass
spectrum is obtained, using as input the Kpi mass spectrum and angular distribution
derived directly from data, without requiring a theoretical description of resonance
shapes or their interference. The hypothesis that the psi(2S)pi mass spectrum can be
described in terms of Kpi reflections alone is rejected with more than 8sigma significance.
This provides confirmation, in a model-independent way, of the need for an additional
resonant component in the mass region of the Z(4430)- exotic state.}
}}

\begin{document}

\renewcommand{\thefootnote}{\fnsymbol{footnote}}
\setcounter{footnote}{1}



\begin{titlepage}
\pagenumbering{roman}

\vspace*{-1.5cm}
\centerline{\large EUROPEAN ORGANIZATION FOR NUCLEAR RESEARCH (CERN)}
\vspace*{1.5cm}
\noindent
\begin{tabular*}{\linewidth}{lc@{\extracolsep{\fill}}r@{\extracolsep{0pt}}}
\ifthenelse{\boolean{pdflatex}}
{\vspace*{-2.7cm}\mbox{\!\!\!\includegraphics[width=.14\textwidth]{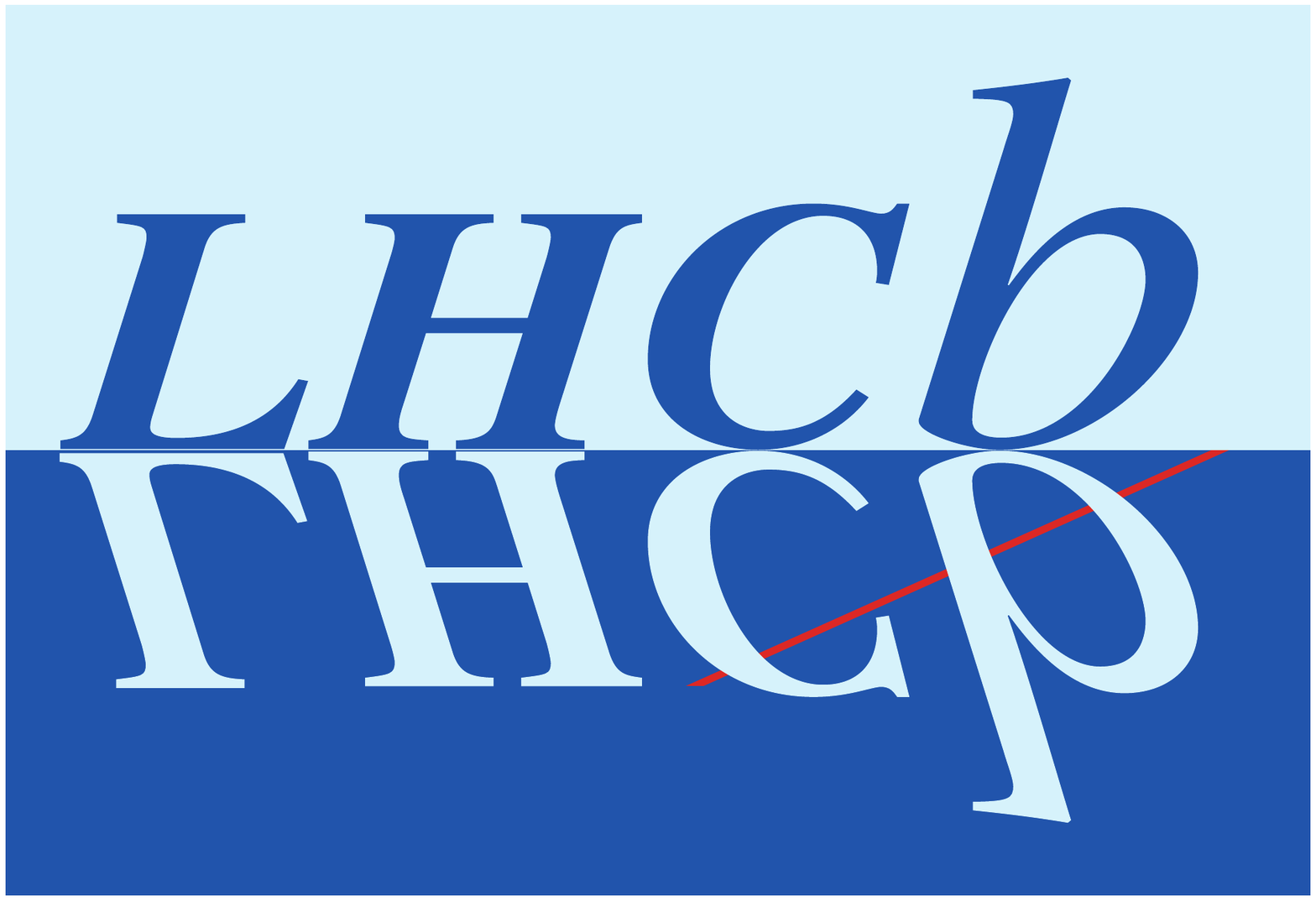}} & &}%
{\vspace*{-1.2cm}\mbox{\!\!\!\includegraphics[width=.12\textwidth]{lhcb-logo.eps}} & &}%
\\
 & & CERN-PH-EP-2015-244 \\  
 & & LHCb-PAPER-2015-038 \\  
 & & 7 October 2015\\ 
 & & \\
\end{tabular*}

\vspace*{4.0cm}

{\bf\boldmath\huge
\begin{center}
  A model-independent confirmation of the $Z(4430)^-$ state
\end{center}
}

\vspace*{2.0cm}

\begin{center}
The LHCb collaboration\footnote{Authors are listed at the end of this paper.}
\end{center}

\vspace{\fill}

\begin{abstract}
  \noindent
	The decay $B^0\to \psi(2S) K^+\pi^-$ is analyzed using $\rm 3~fb^{-1}$ of $pp$ collision data collected with the LHCb detector. 
A model-independent description of the $\psi(2S) \pi$ mass spectrum is obtained, using as input the $K\pi$ mass spectrum and angular distribution derived directly from data, without requiring a theoretical description of resonance shapes or their interference.
The hypothesis that the $\psi(2S)\pi$ mass spectrum can be described in terms of $K\pi$ reflections alone is rejected with more than 8$\sigma$ significance.
This provides confirmation, in a model-independent way, of the need for an additional resonant component in the mass region of the $Z(4430)^-$ exotic state.
\end{abstract}

\vspace*{2.0cm}

\begin{center}
  Published in Phys.Rev. D92, 112009 (2015)  
\end{center}

\vspace{\fill}

{\footnotesize 
\centerline{\copyright~CERN on behalf of the \lhcb collaboration, licence \href{http://creativecommons.org/licenses/by/4.0/}{CC-BY-4.0}.}}
\vspace*{2mm}

\end{titlepage}


\newpage
\setcounter{page}{2}
\mbox{~}
%
%
%
%

\cleardoublepage



\renewcommand{\thefootnote}{\arabic{footnote}}
\setcounter{footnote}{0}
\renewcommand{\topfraction}{0.85}
\renewcommand{\bottomfraction}{0.85}
\renewcommand{\textfraction}{0.1}
\renewcommand{\floatpagefraction}{0.85}


\pagestyle{plain} 
\setcounter{page}{1}
\pagenumbering{arabic}


\newboolean{prl}
\setboolean{prl}{false}

\newlength{\figsize}
\setlength{\figsize}{0.8\hsize}

%
%
\def\BR{{\cal B}}
\def\PDF{{\cal P}}
\def\zff{\ensuremath{Z(4430)^-}\xspace}
\def\zone{\ensuremath{Z_1^-}\xspace}
\def\ME{{\cal M}}
\def\sig{\sigma}
\def\L{{\cal L}}
\def\sWeights{{\mbox{\em sWeights}}}
\def\psip{{\ensuremath{\Ppsi{(2S)}}}\xspace}
\def\psipp{\psip\pi}
\def\mpsipp{m_{\psip\pi}}
\def\mpsipkp{m_{\psipK^+\pi^-}}
\def\mkp{m_{K^+\pi^-}}
\def\Mkp{m_{K\pi^-}}
\def\mmm{m_{\mu^+\mu^-}}
\def\mchicp{m_{\chi_{c1,2}\pi^-}}
\def\MZ{M_{Z_1^-}}
\def\GZ{\Gamma_{Z_1^-}}
\def\ks{K^{*0}}
\def\thetaks{\theta_{K^{*0}}}
\def\cosks{\cos\thetaks}
\def\DLL{D\mathcal{LL}}
\def\thetapsi{\theta_{\psip}}
\def\cospsi{\cos\thetapsi}
\def\ndf{{\rm ndf}}
\def\DeltaL{\Delta(-2\ln L)}

\def\CL{\ensuremath{p_{\chi^2}}\xspace}
\def \costhetaY{\ensuremath{\mathrm{cos(\theta_{\PK\psigen})}}\xspace}   
\def \costhetaZ{\ensuremath{\mathrm{cos(\theta_{\pi\psigen})}}\xspace}   

\def \costheta{\ensuremath{\mathrm{cos({\theta}_{\PK\pi})}}\xspace}   
\def \costhetaX{\ensuremath{\mathrm{cos({\theta}_{\PX(3872)})}}\xspace} 
\def \costhetaPsiGeni{\ensuremath{\mathrm{cos{\theta}^{\it i}_{\psip}}}\xspace}
\def \costhetaPsiGen{\ensuremath{\mathrm{cos{\theta}_{\psip}}}\xspace}
\def \costhetaPsi{\ensuremath{\mathrm{cos({\theta}_{\psitwos})}}\xspace} 
\def \costhetaJpsi{\ensuremath{\mathrm{cos({\theta}_{\jpsi})}}\xspace} 
\def \deltaphi{ \ensuremath{\mathrm{\Delta\phi_{\it{\kaon}\pi,\mu\mu}}}\xspace }
\def \deltaphii{ \ensuremath{\mathrm{\Delta\phi^{\it i}_{\PK\pi,\mu\mu}}}\xspace }
\def \mkpi{\ensuremath{\Pm_{\kaon\pi}}\xspace}
\def \mkpii{\ensuremath{\Pm^i_{\kaon\pi}}\xspace}
\def \mpipi{\ensuremath{\Pm_{\pi\pi}}\xspace}

\def \mpsipigen{\ensuremath{\Pm_{\psigen\pi}}\xspace}
\def \mpsipi{\ensuremath{\Pm_{\psitwos\pi}}\xspace}
\def \mjpsipi{\ensuremath{\Pm_{\jpsi\pi}}\xspace}

\def \mpsiKgen{\ensuremath{\Pm_{\psigen\PK}}\xspace}
\def \mpsiK{\ensuremath{\Pm_{\psitwos\PK}}\xspace}
\def \mjpsiK{\ensuremath{\Pm_{\jpsi\PK}}\xspace}

\def \pjnm{\ensuremath{\langle P_j^N\rangle}\xspace}
\def \pjn{\ensuremath{\langle P_j^N \rangle}\xspace}
\def \pjua #1{\ensuremath{\langle P_{#1}^U\rangle}\xspace}
\def \pjna #1{\ensuremath{\langle P_{#1}^N\rangle}\xspace}
\def \pju{\ensuremath{\langle P_j^U \rangle}\xspace}

\def\kpi        {\ensuremath{\PK\Ppi}\xspace}
\def \Kpi{\ensuremath{\kaon\pion}\xspace}
\def \psipi{\ensuremath{\psitwos\pi}\xspace}
\def \psipigen{\ensuremath{\psigen\pi}\xspace}
\def \jpsipi{\ensuremath{\jpsi\pi}\xspace}
\def \psiK{\ensuremath{\psitwos\PK}\xspace}
\def \psiKgen{\ensuremath{\psigen\PK}\xspace}
\def \jpsiK{\ensuremath{\jpsi\PK}\xspace}

\def \BzToZK{\ensuremath{\Bz \to  \Kp \PZ(4430)^{-}}\xspace}
\def \BpToKPiPiJpsi{\ensuremath{\Bp \to  \Kp \pip\pim \jpsi}\xspace}
\def \BpToXK{\ensuremath{\Bp \to  \Kp \X}\xspace}
\def \BpToPsiK{\ensuremath{\Bp \to  \Kp \psitwos}\xspace}
\def \BzToKPiPsi{\ensuremath{\Bz \to \psip \Kp \pim }\xspace}
\def \BzToKPiJpsi{\ensuremath{\Bz \to  \Kp \pim \jpsi}\xspace}
\def \BzToKPiPsiGen{\ensuremath{\Bz \to  \Kp \pim \psigen}\xspace}
\def \XToJpsipipi{\ensuremath{\PX(3872)\to\jpsi\pipi}\xspace}
\def \Kstthree{\ensuremath{\PK^{\ast}_3(1780)^0}\xspace }
\def \Kstfour{\ensuremath{\PK^{\ast}_4(2045)^0}\xspace }
\def \Kstfive{\ensuremath{\PK^{\ast}_5(2380)^0}\xspace }
\def \Kstz{\ensuremath{\PK^{\ast 0}}\xspace }
\def \Kst{\ensuremath{\PK^{\ast}}\xspace }
\def \Kappa{\ensuremath{\PK^{\ast}(800)^0}\xspace }
\def \Kstuno{\ensuremath{\PK^{\ast}(892)^0}\xspace }
\def \Kstdue{\ensuremath{\PK^{\ast}_2(1430)^0}\xspace }
\def \Kstzero{\ensuremath{\PK^{\ast}_0(1430)^0}\xspace }
\def \KstMQ{\ensuremath{\PK^{\ast}(1410)^0}\xspace }
\def \KstMS{\ensuremath{\PK^{\ast}(1680)^0}\xspace }
\def \Zp{\ensuremath{\PZ(4430)^{-}}\xspace }
\def \ZpToPsiPi{\ensuremath{\PZ(4430)^{-}\to\psitwos\pim}\xspace }
\def \dimuon{\ensuremath{\text{di-}\mmu}\xspace }
\def \X{\ensuremath{\PX(3872)}\xspace}
\def \jpc{\ensuremath{\PJ^{PC}}\xspace}
\def \jp{\ensuremath{\PJ^{P}}\xspace}
\def \lmax{ \ensuremath{l_{\text{max}} }\xspace}
\def\psigen  {\ensuremath{\Ppsi}\xspace}
\def\Psigma      {\ensuremath{\upsigma}\xspace}                 
\def \deltanll{\ensuremath{\Delta NLL}\xspace}
\newcommand{\NLL}{\ensuremath{\Delta\textrm{NLL}}}
\def \brown{\color{brown}}
\def \blue{\color{blue}}
\def \green{\color{green}}
\def \red{\color{red}}
\def \magenta{\color{magenta}}
\def \costhetaformula{\ensuremath{\mathrm{cos(\theta_{\PK})} =
\frac{(p_{\Bz}\cdot p_{\PK})\mkpi^2 - (p_{\Bz}\cdot
p_{\kaon\pion})(p_{\kaon\pion}\cdot p_{\PK})}{\sqrt{\left[ (p_{\Bz}\cdot
p_{\kaon\pion})^2 - m_{\Bz}^2\mkpi^2\right] \left[ (p_{\kaon\pion}\cdot
p_{\kaon})^2 - \mkpi^2 m_{\kaon}^2\right] }} }\xspace}

\def\tempclearpage{\clearpage}

\mathchardef\myhyphen="2D


\let\clearpage\relax
\section{Introduction}
\label{sec:introduction}

\noindent

Almost all known mesons and baryons can be described in the quark model with 
combinations of two or three quarks, although the existence of higher multiplicity 
configurations, as well as additional gluonic components, is, in principle, not excluded \cite{GellMann:1964nj}.
For many years significant effort has been devoted to the search 
for such exotic configurations.
In the baryon sector, resonances 
with a five-quark content have been searched for extensively \cite{Bevan:2014iga,Ireland:2007aa,Hicks:2012zz,PDG2014}. Recently, LHCb  has observed a resonance in the $\jpsi p$ channel, compatible with being a pentaquark-charmonium state \cite{LHCb-PAPER-2015-029}.
In the meson sector, several charmonium-like states, that could be interpreted as four-quark states 
\cite{Drenska:2010kg,Olsen:2014qna}, have been reported by a number of
experiments but not all of them have been confirmed.
\par The existence of the \Zp hadron, originally observed by the Belle collaboration \cite{Choi:2007wga,Mizuk:2009da,Chilikin:2013tch} in the decay $\BzToZK$ with $\ZpToPsiPi$, was confirmed by the LHCb collaboration \cite{LHCb-PAPER-2014-014} (the inclusion of charge-conjugate processes is implied). 
This state, having a minimum quark content of $c\bar c u\bar d$, is the strongest candidate for a four-quark meson \cite{Rosner:2007mu,Braaten:2007xw,Cheung:2007wf,Li:2007bh,Qiao:2007ce,Liu:2008qx,Maiani:2008zz,Bugg:2008wu,Matsuki:2008gz,Branz:2010sh,Galata:2011bi,Nielsen:2014mva}.
Through a multidimensional amplitude fit LHCb confirmed
the existence of the \Zp resonance with a significance of $13.9\sigma$ and its mass and width were measured to be  $M_{Z^-}=4475\pm 7\,{^{+15}_{-25}}\mevcc$ and $\Gamma_{Z^-}=172\pm 13{^{+37}_{-34}}\mevcc$. 
Spin-parity of $J^P=1^+$ was favoured over the other assignments by more than $17.8\sigma$ and,  
through the study of the variation of the phase of the $\Zp$ with mass, LHCb demonstrated its resonant character.
\par The BaBar collaboration \cite{Aubert:2008aa} searched for the $\Zp$ state 
in a data sample statistically comparable to Belle's. 
They used a model-independent approach to test whether an interpretation of the experimental data is possible 
in terms of the known resonances in the \Kpi system. The \Kpi mass and angular distributions
were determined from data and used to predict the observed  $\psipp$ 
mass spectrum. It was found that
the observed  $\psipp$  mass spectrum was compatible with being described by reflections of $\Kpi$ system.  
Therefore no clear evidence for a $\Zp$ was established, although BaBar's analysis did not exclude the observation by Belle.
\par The present article describes the details of an LHCb analysis that was briefly reported in Ref. \cite{LHCb-PAPER-2014-014}.
Adopting a model-independent approach, along the lines of BaBar's strategy, the structures observed in the 
$\psip \pi$ mass spectrum are predicted in terms of the reflections of the \Kpi system mass and angular composition, without introducing any modelling of the resonance lineshapes and their interference patterns.  The compatibility of these predictions with data is quantified.



\ifthenelse{\boolean{prl}}{ }
\section{The LHCb detector}
\label{sec:detector}


The \lhcb detector\cite{Alves:2008zz,LHCb-DP-2014-002} is a single-arm forward
spectrometer covering the \mbox{pseudorapidity} range $2<\eta <5$,
designed for the study of particles containing \bquark or \cquark
quarks. The detector includes a high-precision tracking system
consisting of a silicon-strip vertex detector surrounding the $pp$
interaction region \cite{LHCb-DP-2014-001}, a large-area silicon-strip detector located
upstream of a dipole magnet with a bending power of about
$4{\rm\,Tm}$, and three stations of silicon-strip detectors and straw
drift tubes~\cite{LHCb-DP-2013-003}  placed downstream of the magnet.
The tracking system provides a measurement of momentum, \ptot, of charged particles with
a relative uncertainty that varies from 0.5\% at low momentum to 1.0\% at 200\gevc.
The minimum distance of a track to a primary vertex, the impact parameter, is measured with a resolution of $(15+29/\pt)\mum$,
where \pt is the component of the momentum transverse to the beam, in\,\gevc.
Different types of charged hadrons are distinguished using information
from two ring-imaging Cherenkov detectors~\cite{LHCb-DP-2012-003}. 
Photons, electrons and hadrons are identified by a calorimeter system consisting of
scintillating-pad and preshower detectors, an electromagnetic
calorimeter and a hadronic calorimeter. Muons are identified by a
system composed of alternating layers of iron and multiwire
proportional chambers~\cite{LHCb-DP-2012-002}.
The online event selection is performed by a trigger~\cite{LHCb-DP-2012-004}, 
which consists of a hardware stage, based on information from the calorimeter and muon
systems, followed by a software stage, which applies a full event
reconstruction.



\ifthenelse{\boolean{prl}}{ }
\section{Data samples and candidate selection}
\label{sec:selection}

\begin{figure}[t]
\centering
\includegraphics[width=.5\linewidth]{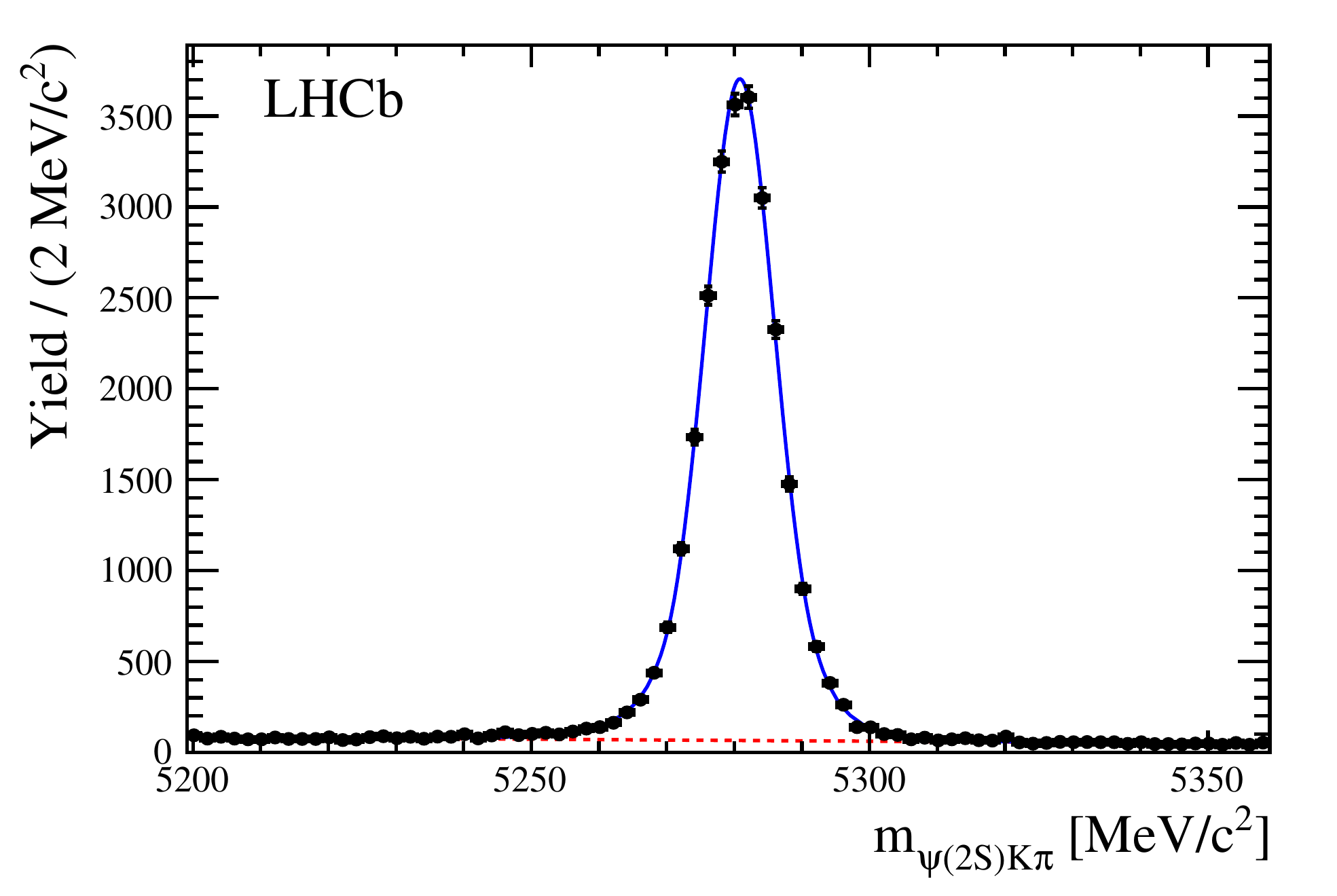}
\caption{Spectrum of the $\psip \kaon \pi$ system invariant mass. Black dots are the data, the continuous (blue) line represents the fit result and the dashed (red) line represents the background component.} 
\label{fig:bmassspectrum}
\end{figure}

The results presented in this paper are based on data from ${\it pp}$ collisions collected by the LHCb experiment, corresponding to integrated luminosities of $1\invfb$ and $2\invfb$ at center-of-mass energies of $7\tev$ in 2011 and $8\tev$ in 2012, respectively.

In the simulation, ${\it pp}$ collisions are generated using PYTHIA \cite{Sjostrand:2006za,*Sjostrand:2007gs} with a specific LHCb configuration \cite{LHCb-PROC-2010-056}. The decays of hadronic particles are described by EvtGen \cite{Lange:2001uf}, in which
final-state radiation is generated using PHOTOS \cite{Golonka:2005pn}. 
The interaction of the generated
particles with the detector, and its response, are implemented using the GEANT4 toolkit \cite{Allison:2006ve,*Agostinelli:2002hh} as described in Ref. \cite{LHCb-PROC-2011-006}.
Samples of simulated events, generated with both 2011 and 2012 conditions, are produced for the decay $\BzToKPiPsi$ with a uniform 3-body phase-space distribution and the $\psip$ decaying into two muons. These simulated events are used to tune the event selection and for efficiency and  resolution studies. 
\par The selection is similar to that used in Ref. \cite{LHCb-PAPER-2014-014} and consists of a cut-based preselection followed by a multivariate analysis. Track-fit quality and particle identification requirements are applied to all charged tracks. 
The $\Bz$ candidate reconstruction starts by requiring two well-identified muons, with opposite charges, 
having  $\pt>2\gevc$ and forming a good quality vertex.
The dimuon invariant mass has to lie in the window 3630--3734$\mevcc$, around the $\psip$ mass.
To obtain a $\Bz$ candidate, each dimuon pair is required to form a good vertex with a kaon 
and a pion candidate, with opposite charges.
Pions and kaons are required to be inconsistent with coming from any primary vertex (PV) and to have  tranverse momenta
greater than $200\mevc$.  
    
The $\Bz$ candidate has to have $\pt>2\gevc$, a reconstructed decay time exceeding 
0.25$\ps$ and an invariant mass in the window 5200--5380$\mevcc$ around the nominal $\Bz$ mass.
Contributions from $\phi \rightarrow K^+K^-$ decay, where one of the kaons is misidentified as a
pion, are removed by vetoing the region 1010--1030$\mevcc$ of the dihadron
invariant mass calculated assuming that the \pim candidate has the \Km mass.
\par To reduce the combinatorial background, a requirement is imposed on the output of a multivariate
discriminator based on the likelihood ratio \cite{Hocker:2007ht}.

The four variables used as input are: the smaller $\chi^2_{IP}$ of the kaon and the pion, where $\chi^2_{IP}$ is the
difference in the PV fit $\chi^2$ with and without the track under consideration; the $\mu^+\mu^-\kaon\pion$
vertex-fit quality;
the \Bz candidate impact parameter significance with respect to the PV; and the cosine of the largest opening 
angle between the $\psip$ and each of the charged hadrons in the plane transverse to the beam.
After the multivariate selection, the $\Bz$ candidate invariant-mass distribution appears as shown in \autoref{fig:bmassspectrum} with a fitted curve superimposed. 
The fit model consists of a Hypatia distribution\cite{Santos:2013gra} to describe the signal, and an exponential function to describe the background.

~\autoref{tab:bfit} provides the fit results and the signal and background yields in the signal region. The width of the distribution, $\sigma_{\Bz}$, is defined as half the symmetric interval around $M_{\Bz}$ containing 68.7\% of the total signal. The signal region is defined by the $\pm 2\sigma_{\Bz}$ interval around $M_{\Bz}$.

Sideband subtraction is used to remove the background which is dominated by combinations of $\psip$ mesons from \bquark-hadron decays with random kaons and pions. 
Sidebands are identified by the intervals $[M_{\Bz}-80$, $M_{\Bz}-7\sigma_{\Bz}]$\mevcc and $[M_{\Bz}+7\sigma_{\Bz}$, $M_{\Bz}+80]$\mevcc. A weight, $W_{\text{signal}}$,
is attributed to each candidate: unit weight is assigned to candidates in the signal region; the ratio of the background yield in the signal region and in the sidebands, with a negative sign, is the weight assigned to sideband candidates; zero weight is assigned to candidates in the remaining regions.

  \begin{table}[t]
  \begin{center}
  \begin{tabular}{lr@{$\pm$}l} 
   {Variable}  &  \multicolumn{2}{c}{Fit results} \\
   \hline
   $M_{\Bz}$                 &\phantom{0}5280.83 & 0.04 \mevcc\\
   $\sigma_{\Bz}$            &\phantom{0}5.77    & 0.05 \mevcc\\
   Signal yield              &\phantom{0}23,801  & 158\\
   Background yield          &\phantom{0}757     & 14\\       
    \end{tabular}
    \end{center}
    \caption{ Results of the fit to the invariant mass spectrum of the $\psip \kaon \pi$ system. 
    The signal and the background yields are calculated in the signal region defined by the interval of $\pm 2\sigma_{\Bz}$ around $M_{\Bz}$.}
    \label{tab:bfit}
  \end{table}
  
  \FloatBarrier



\ifthenelse{\boolean{prl}}{ }{ \tempclearpage }

\section{Efficiency and resolution}
\label{sec:effres}

   \begin{figure}[H]
   \centering
   \begin{subfigure}[b]{.330\linewidth}
   \includegraphics[width=\linewidth]{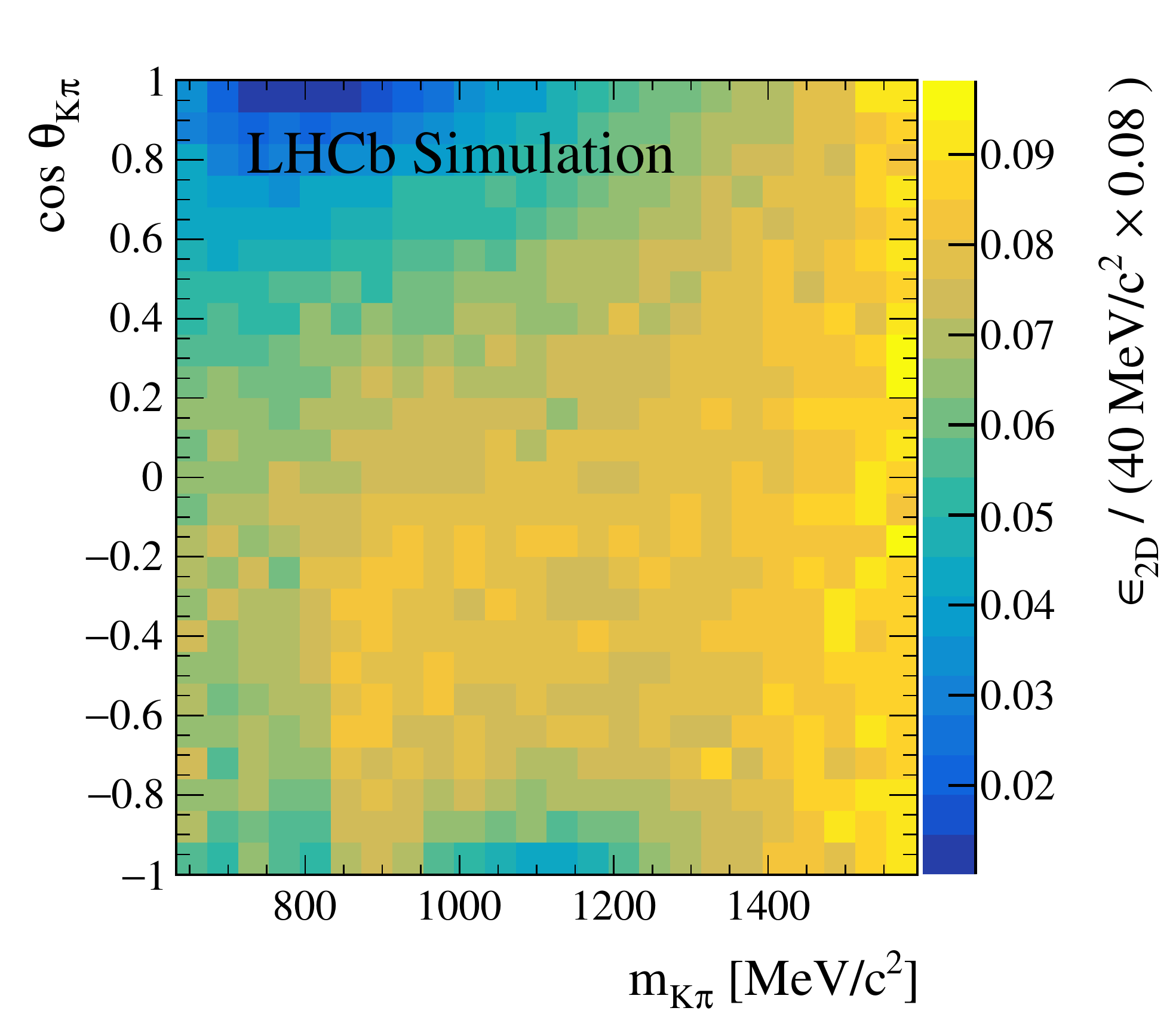}
   \end{subfigure}
   \begin{subfigure}[b]{.330\linewidth}
    \includegraphics[width=\linewidth]{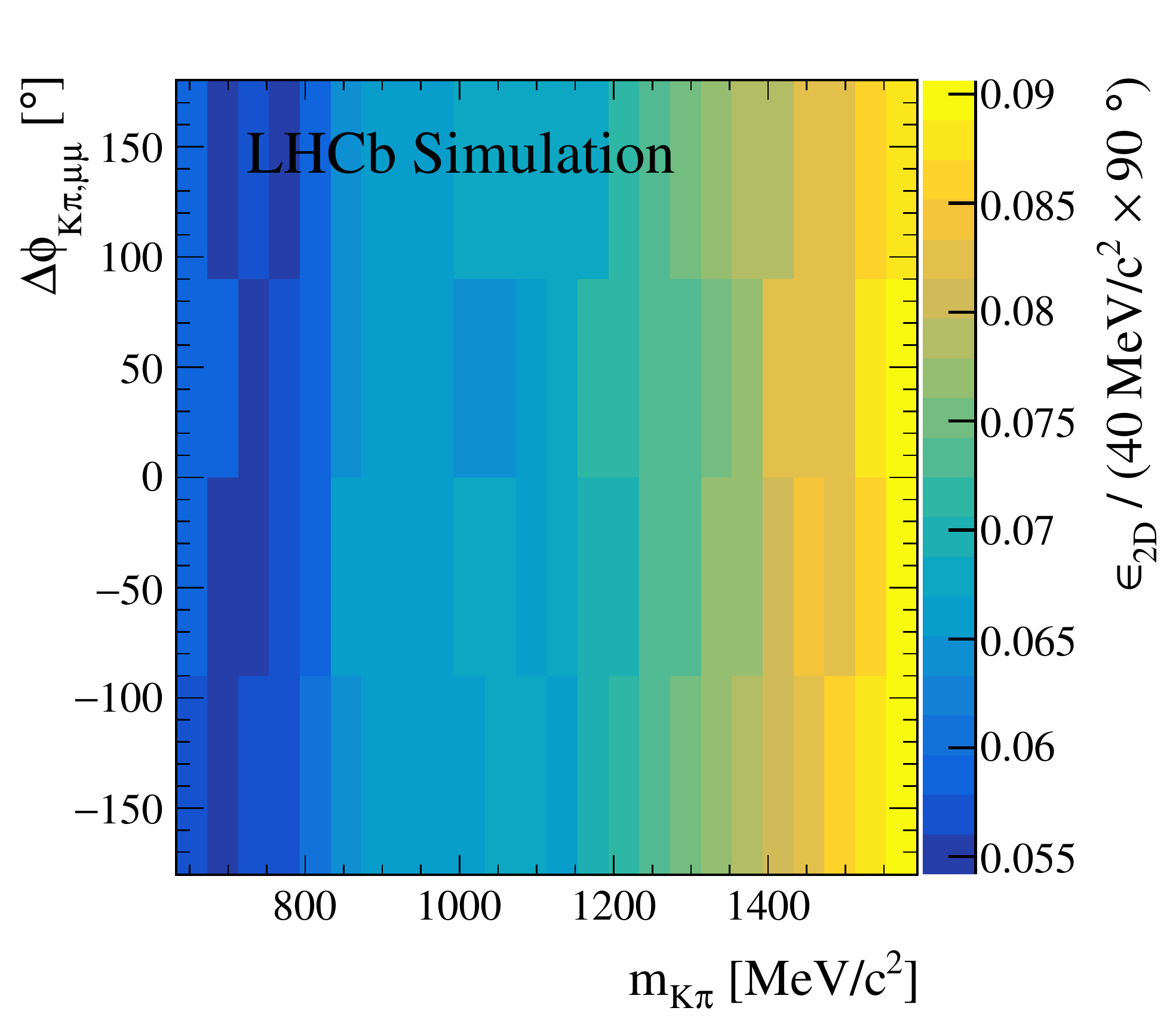}
   \end{subfigure}
   \begin{subfigure}[b]{.330\linewidth}
   \includegraphics[width=\linewidth]{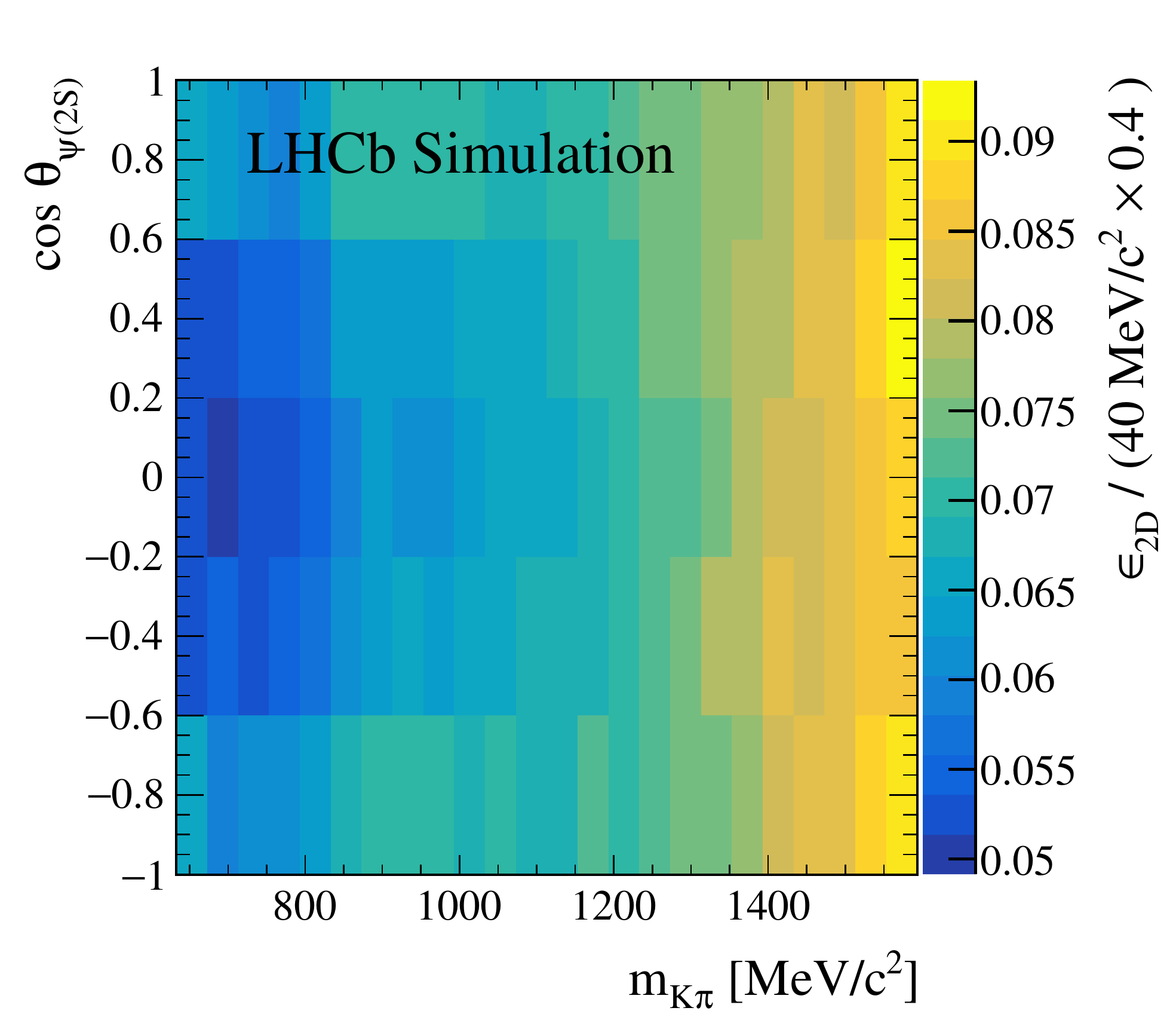}
   \end{subfigure}
	\\
	   \begin{subfigure}[b]{.330\linewidth}
	 \includegraphics[width=\linewidth]{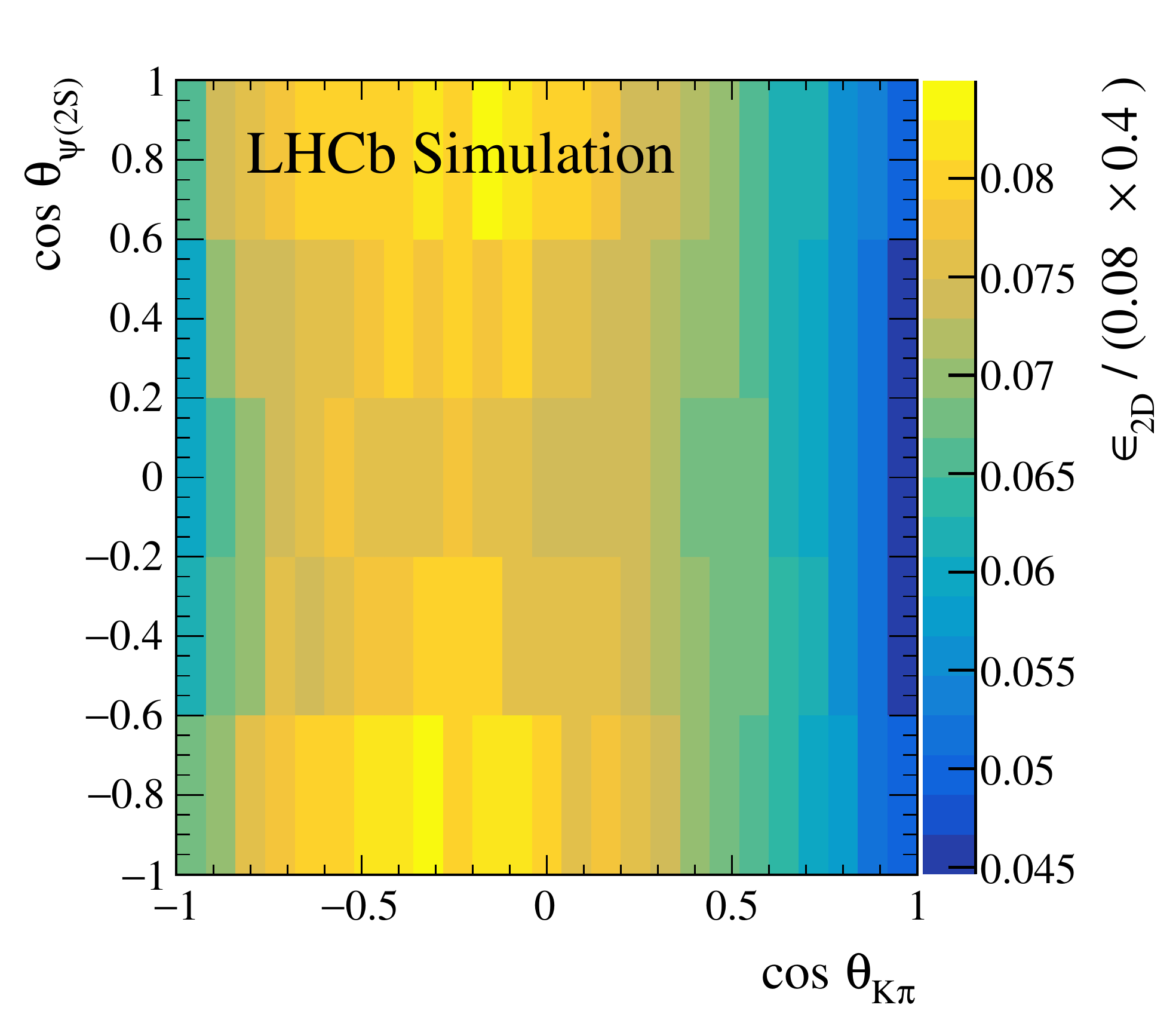}
   \end{subfigure}
   \begin{subfigure}[b]{.330\linewidth}
   \includegraphics[width=\linewidth]{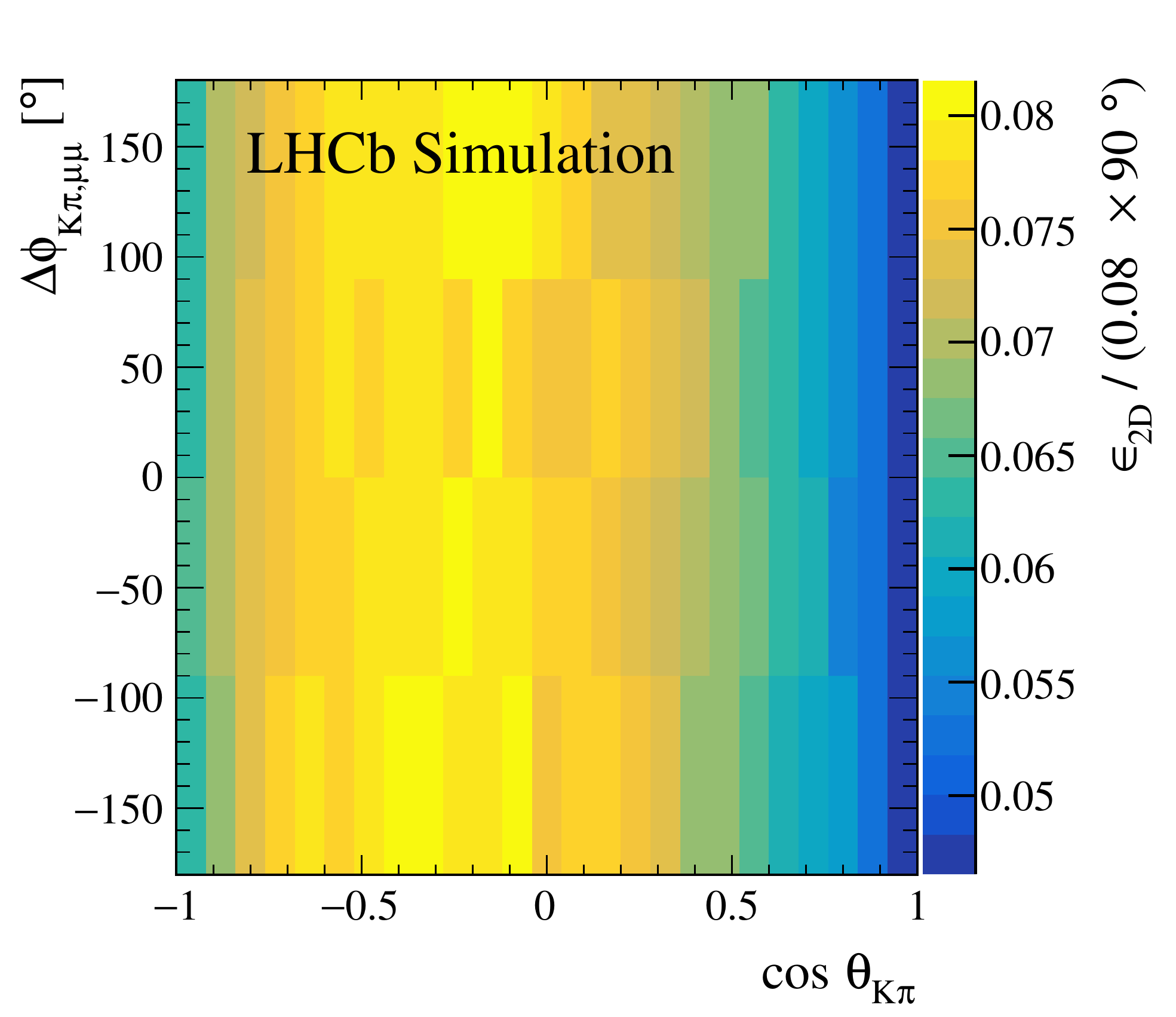}
   \end{subfigure}
   \begin{subfigure}[b]{.330\linewidth}
   \includegraphics[width=\linewidth]{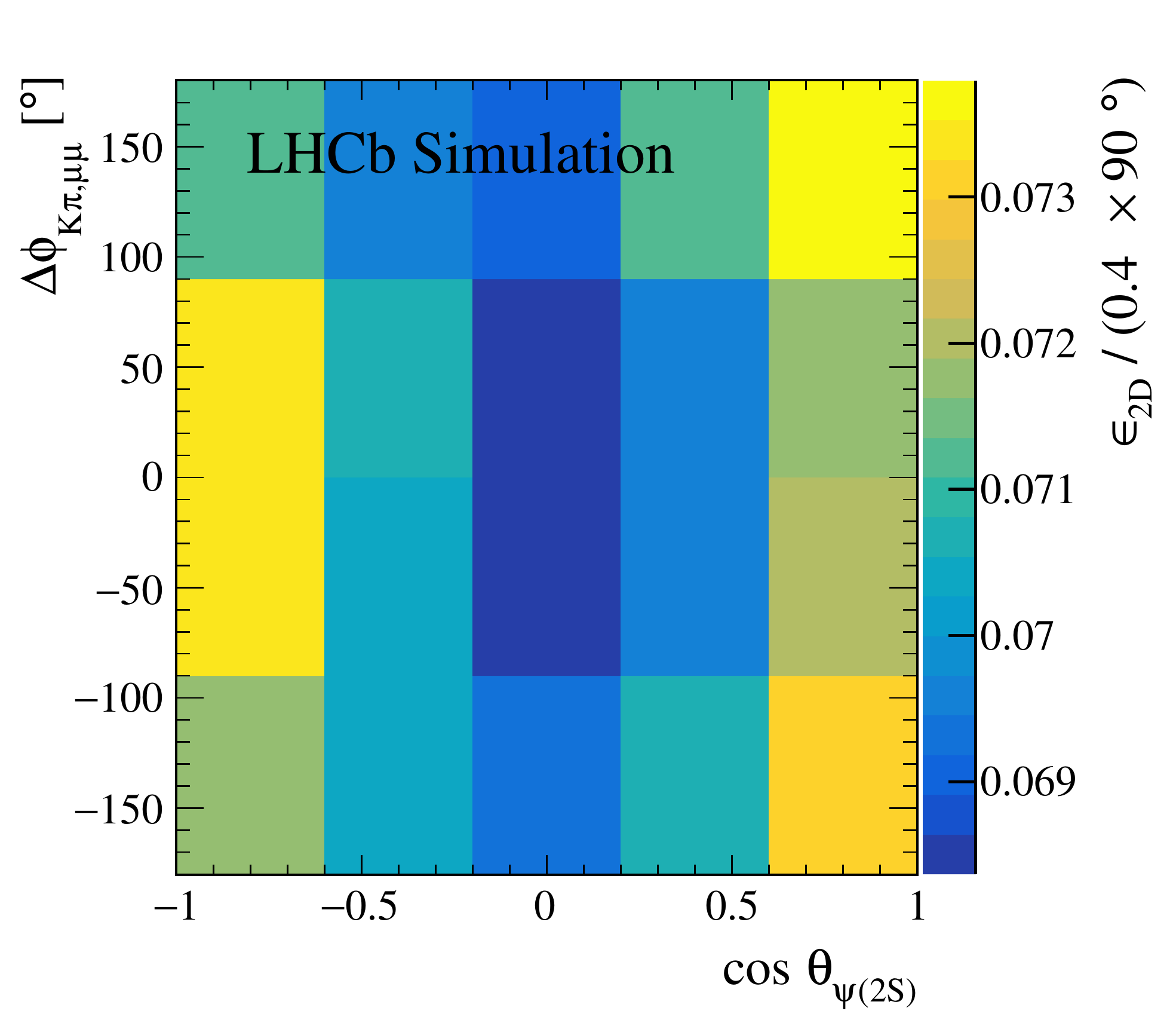}
   \end{subfigure}
	 \caption{\small The 2D efficiency shown in the following planes: 
 				(top-left) ($\mkpi$,$\cosks$), (top-middle) ($\mkpi$,$\deltaphi$), (top-right) ($\mkpi$,$\costhetaPsiGen$), (bottom-left) ($\cosks$,$\costhetaPsiGen$), (bottom-middle) ($\cosks$,$\deltaphi$), (bottom-right) ($\costhetaPsiGen$,$\deltaphi$). 
				Corrections for the efficiency are applied in the 4D space; the 2D plots allow
				visualization of their behavior.}
 \label{fig:eff_2D_4D_psi}
 \end{figure}

\begin{figure}[t]
\centering
\includegraphics[width=0.50\linewidth]%
{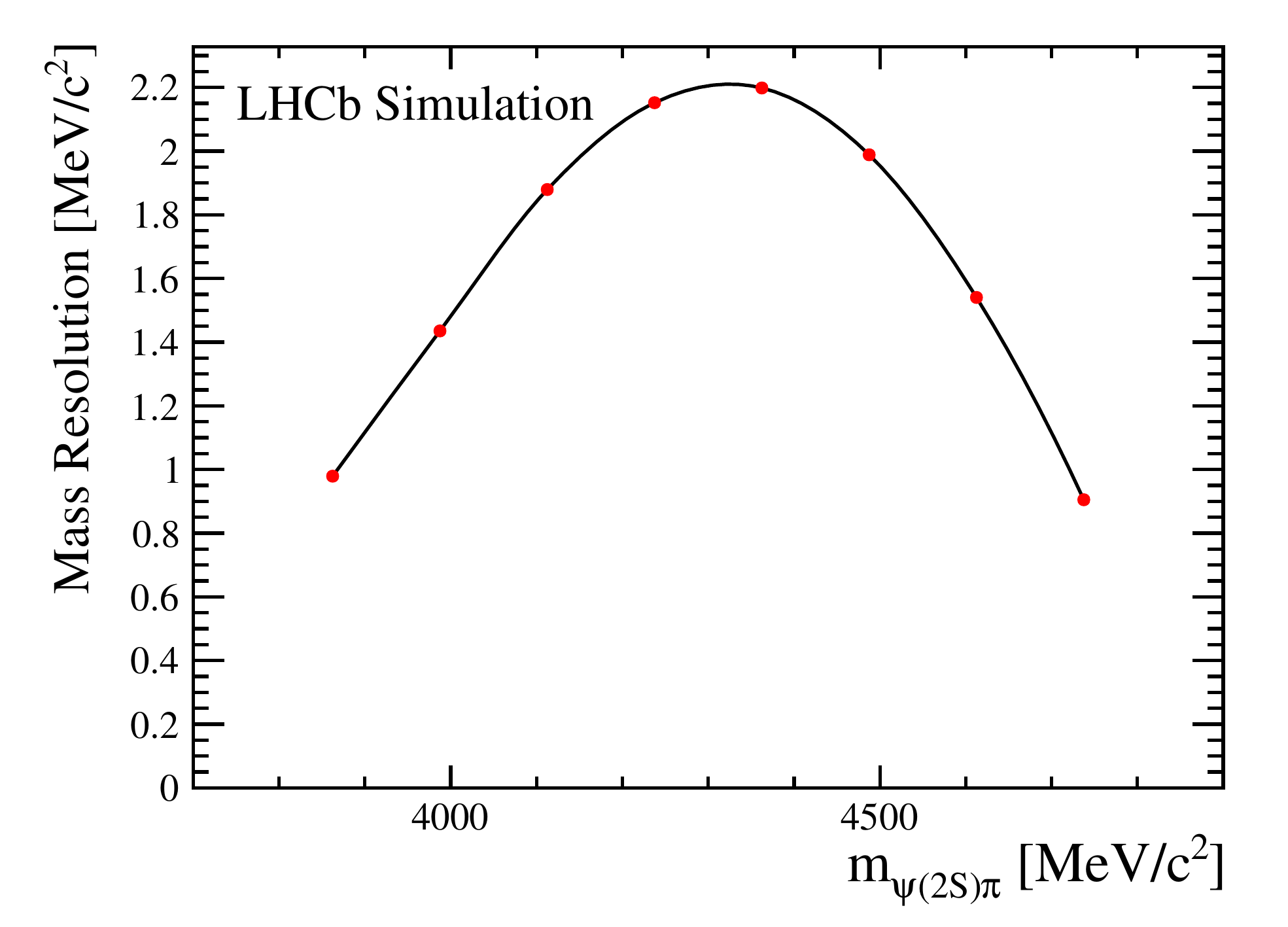}
\caption{\small The $\psipp$ invariant mass resolution as determined from simulated data (red dots). The continuous line is a spline-based interpolation. }
\label{fig:res}
\end{figure}

\begin{table}[b]
 \begin{center}
 \begin{tabular}{@{\extracolsep{\fill}}ll} 
   Variable  & \multicolumn{1}{c}{Resolution}\\ 
   \hline
   $\mkpi$    & \phantom{0}1.5\mevcc\\ 
   $\mpsipp$ &\phantom{0}1.8\mevcc\\    
   $\cosks$      &\phantom{0}0.004\\
   $\costhetaPsiGen$    & \phantom{0}0.005\\
   $\deltaphi$& \phantom{0}0.3$\degrees$ \\           
   \end{tabular}
   \end{center}
   \caption{\small Experimental resolution of kinematical quantities, as estimated from Monte Carlo simulations.}
   \label{tab:average-res}
  \end{table}

\par The reconstruction and selection efficiency has been 
evaluated using simulated samples.
The efficiency is calculated as a function 
of four variables: the $K\pi$ system invariant mass, $\mkpi$;
the cosine of the $\ks$ helicity angle, $\cosks$; the cosine of the $\psip$ helicity angle, $\costhetaPsiGen$; and the angle between the $K\pi$ and the $\mu^+\mu^-$ planes calculated in the $B^0$ rest frame, $\deltaphi$ (this variable is called $\phi$ in Ref.\cite{LHCb-PAPER-2014-014}). The helicity angle of the $\Kstz$ ($\psip$) 
is defined as the angle between the $K^+$ ($\mu^+$) direction
and the \Bz direction in the $\Kstz$ ($\psip$) rest frame.
This 4D space is subdivided in 24, 25, 5 and 4 bins of the respective variables.  The value of the efficiency, at each point of the 4D space, is evaluated as a multilinear interpolation of the values at the sixteen bins centers surrounding it. To the points falling in a border 4D bin, where interpolation is not possible, the value of the efficiency at the bin center is assigned.

To visualize the behavior, 2D
efficiency plots are shown, as functions of all the possible variable pairs, in \autoref{fig:eff_2D_4D_psi}.

\par Table ~\ref{tab:average-res} lists the resolutions (average uncertainty) of the reconstructed event variables as evaluated on simulated events. 
They are found to be very small compared to the width of any possible structure searched for in this analysis; therefore no resolution corrections are applied.
In addition, the smooth behavior of the $\mpsipp$ resolution, shown in \autoref{fig:res}, demonstrates
that structures in the $\mpsipp$ spectrum could not be caused by resolution effects.

\FloatBarrier



\ifthenelse{\boolean{prl}}{ }{ \tempclearpage }
\section{$K^*$ resonances}
\label{sec:kstars}

\begin{figure}[h]
\centering
\begin{subfigure}[b]{.45\linewidth}
\includegraphics[width=\linewidth]%
{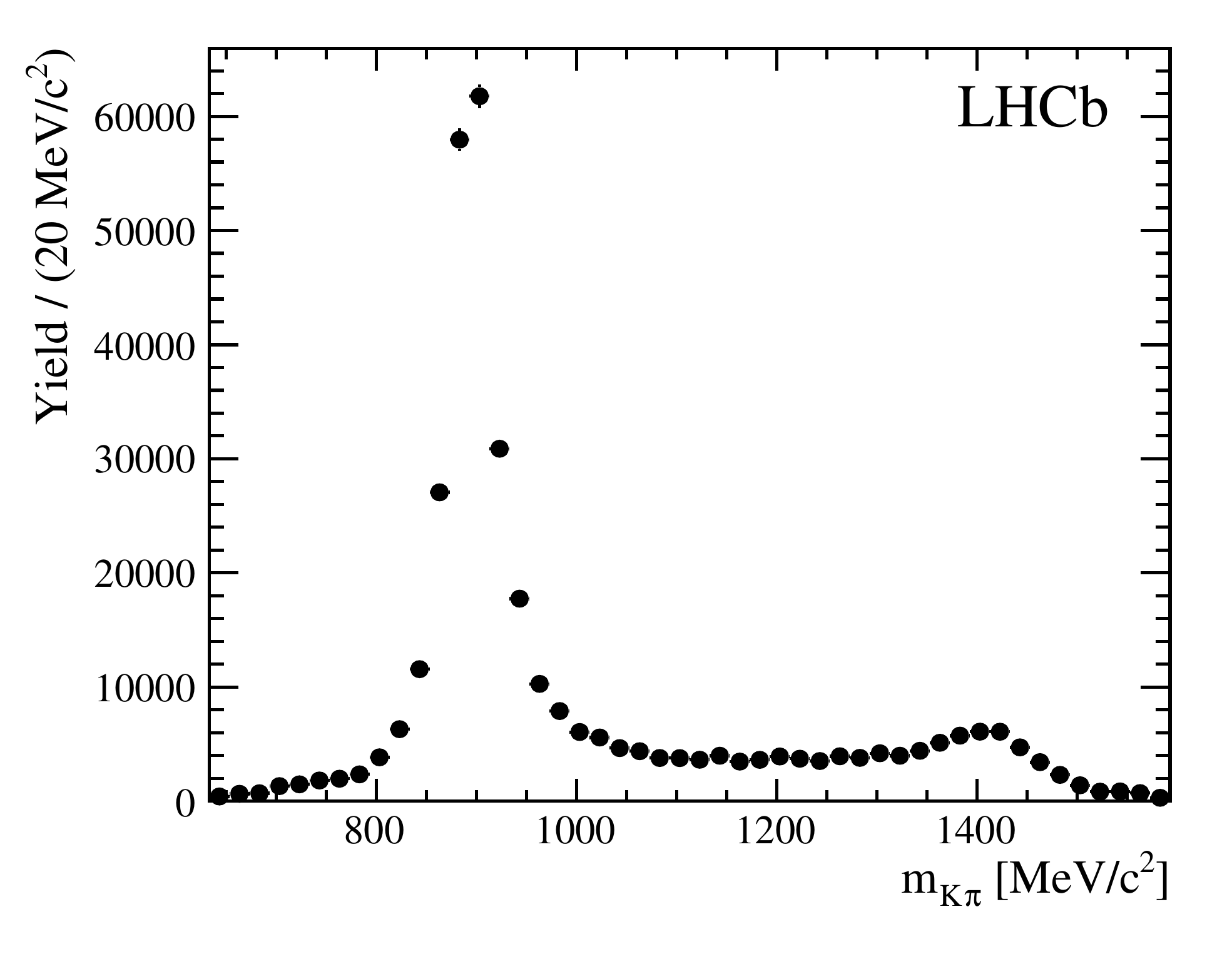}
\end{subfigure}%
\begin{subfigure}[b]{.45\linewidth}
\includegraphics[width=\linewidth]%
{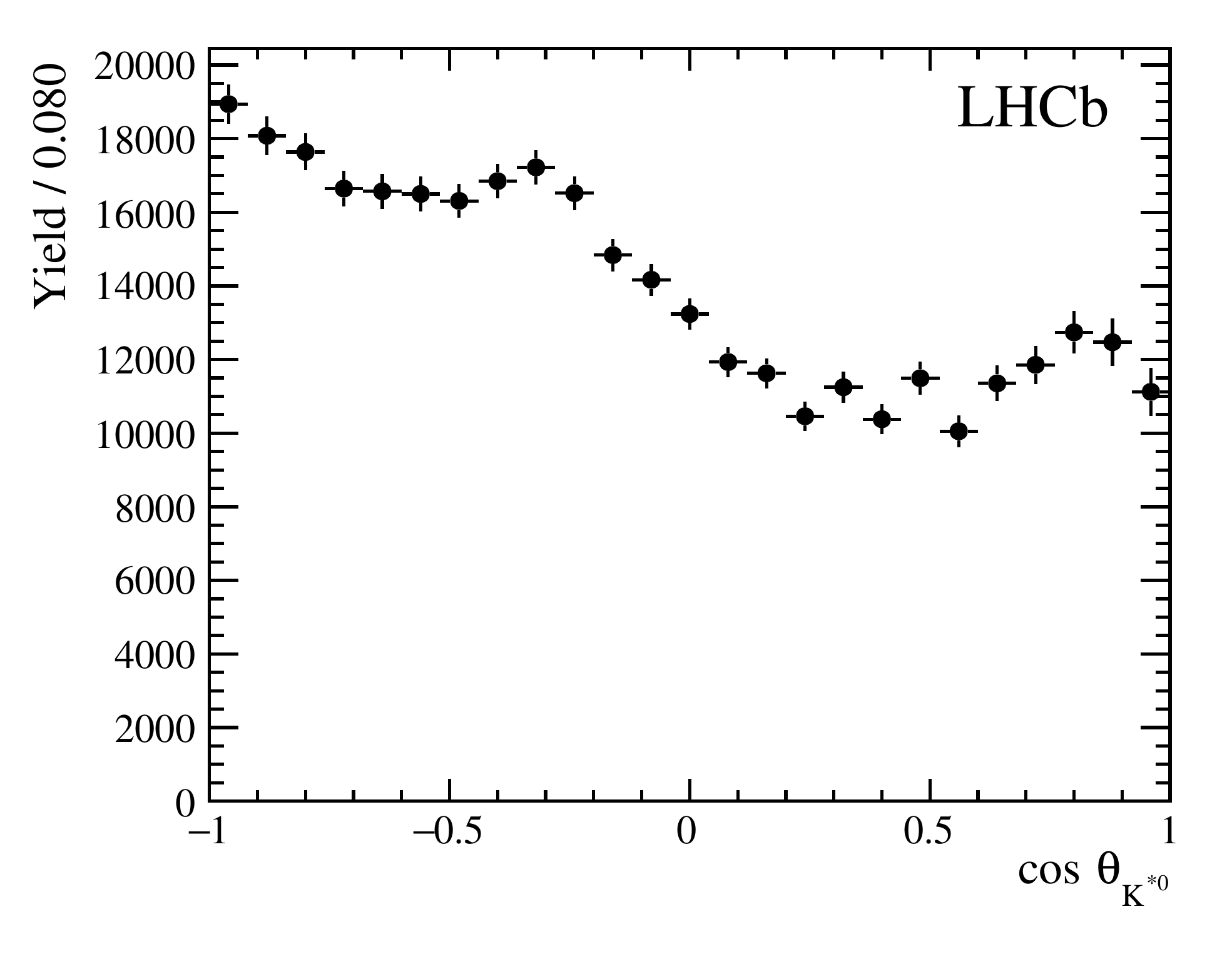}
\end{subfigure}%
\caption{\small Distributions  of (left) \mkpi and (right) $\cosks$,  after background subtraction and efficiency correction.}
\label{fig:kstars_monodim}
\end{figure}

A number  of $K^{*0}$ resonances  with masses up to slightly above the kinematic limit of $1593\mevcc$ can decay to
the \Kpi final state and  contribute to the $\BzToKPiPsi$ decay.
\autoref{tab:kstars} lists these $K^{*0}$ states as well as resonances just above the kinematic limit.

\begin{table}[b] 
  \centering
  \begin{tabular}{@{\extracolsep{\fill}}lr@{$\pm$}lr@{$\pm$}lc} 
     Resonance    & \multicolumn{2}{c}{ Mass (\mevcc)}    & \multicolumn{2}{c}{$\Gamma$ (\mevcc)} & \jp   \\
     \hline
		 \Kappa      & \phantom{0}682    & 29     & \phantom{0}547  & 24  & $0^{+}$\\
     \Kstuno     & \phantom{0}895.81 & 0.19   & \phantom{0}47.4 & 0.6 & $1^{-}$\\
     \KstMQ      & \phantom{0}1414   & 15     & \phantom{0}232  & 21  & $1^{-}$\\
     \Kstzero    & \phantom{0}1425   & 50     & \phantom{0}270  & 80  & $0^+$  \\
     \Kstdue     & \phantom{0}1432.4 & 1.3    & \phantom{0}109  & 5   & $2^+$  \\
     \KstMS      & \phantom{0}1717   & 27     & \phantom{0}322  & 110 & $1^-$  \\
     \Kstthree   & \phantom{0}1776   & 7      & \phantom{0}159  & 21  & $3^-$  \\                          
    \end{tabular}
    \caption{\small Mass, width, spin and parity of resonances known to decay to the \Kpi final state \cite{PDG2014}. The list is limited to masses up to just above the maximum invariant mass for the \Kpi system which, in the decay $\BzToKPiPsi$, is $1593\mevcc$.}
    \label{tab:kstars}
  \end{table}

\begin{figure}[t]
\centering
\begin{subfigure}[b]{.450\linewidth}
\includegraphics[width=\linewidth]%
{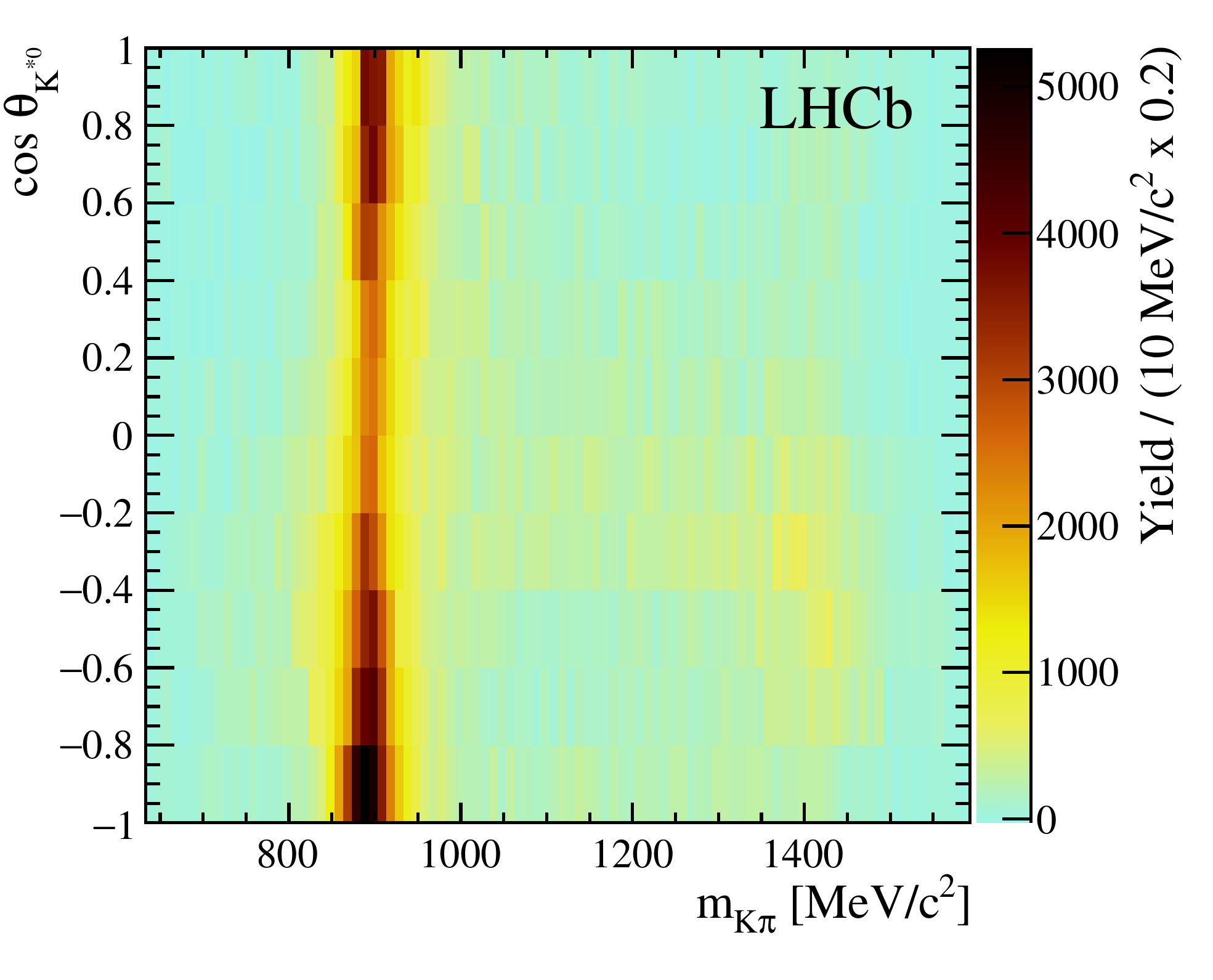}
\end{subfigure}%
\begin{subfigure}[b]{.450\linewidth}
\includegraphics[width=\linewidth]%
{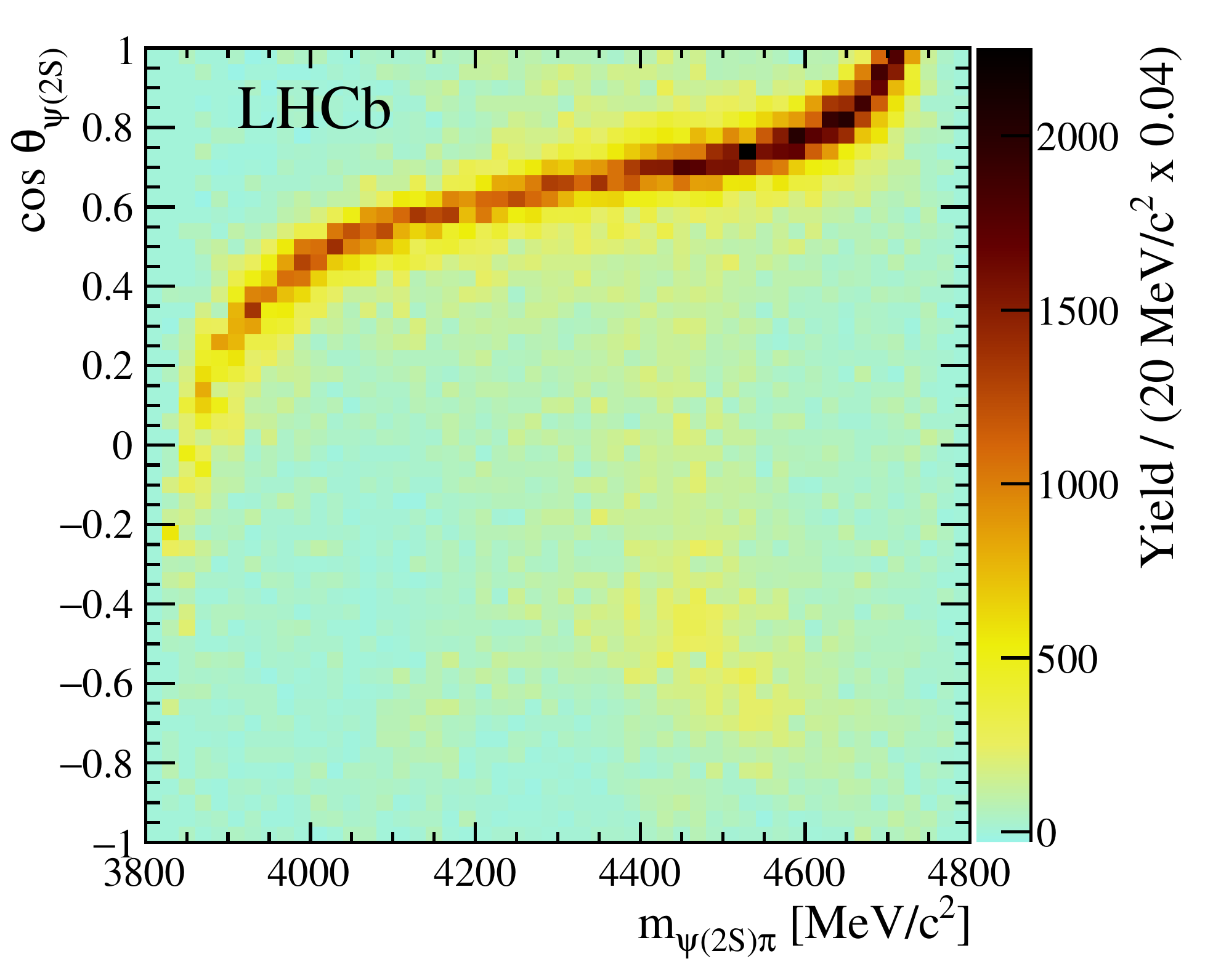}
\end{subfigure}
\caption{\small The two-dimensional distributions (\mkpi,$\cosks$) and ($\mpsipp$,$\costhetaPsiGen$) are shown in the left and the right plots, respectively, after background subtraction and efficiency correction.}
\label{fig:kstars_rect_dalitz}
\end{figure}
  
The \mkpi spectrum of candidate events, shown in the left plot of ~\autoref{fig:kstars_monodim}, is dominated by the $\Kstuno$ meson. A structure in the \KstMQ, $\Kstzero$ and $\Kstdue$ mass region is also clearly visible. In addition, a non-resonant component is evident.
A contribution from the low-mass tail of excited states above the kinematic limit is expected, in particular  from the spin-1  \KstMS and the spin-3 \Kstthree due to their large widths. 
The right plot of \autoref{fig:kstars_monodim} shows the 
  $\cosks$ distribution which highlights the rich angular structure of the $\Kpi$ system.
The resonant structures of the $\Kpi$ system can be also seen in the 2D distributions shown in \autoref{fig:kstars_rect_dalitz}. The plot on the right 
illustrates how the structures present in the \Kpi system considerably influences the $\psipp$ system.

\FloatBarrier

\nopagebreak


\ifthenelse{\boolean{prl}}{ }{ \tempclearpage }
\nopagebreak
\section{Extraction of the moments of the \kpi system}
\label{sec:MIMomentsExtraction}

Background-subtracted and efficiency-corrected data are subdivided in \mkpi bins of width $30\mevcc$, which is suitable for observing the $\Kpi$ resonance structures. 
For each \mkpi bin,
the $\cosks$ distribution can be expressed as an expansion in terms of Legendre polynomials. 
The coefficients of this expansion contain all of the  
information on the angular structure of the system and
characterize the spin of the contributing resonances.
The angular distribution, after integration over the $\psip$ decay angles, can be written as 
\begin{equation}
\frac{dN}{d\cosks}  = \sum_{j=0}^{\lmax}\pju\PP_j(\cosks),
\label{eq:exp_legendre}
\end{equation}

\noindent where
 \lmax depends on the maximum orbital
angular momentum necessary to describe the \kpi system, $\PP_j(\cosks) =
\sqrt{2\pi}Y_j^0(\cosks)$ are Legendre polynomials and $Y_j^0$ are spherical harmonic functions. 
The coefficients \pju in \autoref{eq:exp_legendre} are
called unnormalized moments (moments, in the following) and can be calculated as integrals of the product of the corresponding Legendre polynomial and the $\cosks$ distribution.
Resonances of the \kpi system with spin $s$ can contribute to the moments up to 
\pjua{2s}. Interference between resonances with spin $s_1$ and $s_2$ can contribute
to moments up to \pjua{s_1+s_2}. 

For large samples, the moments are determined from data as
\begin{equation}
\pju = \sum_{i=1}^{N_{\text{reco}}}
\frac{W^i_{\text{signal}}}{\Pepsilon^i}\PP_j(\cosks^i),
\label{eq:unn_moments}
\end{equation}

\noindent where $\PN_{\text{reco}}$ is the number of reconstructed and selected 
candidates in the \mkpi bin. The superscript $i$ labels the candidate, $W^i_{\text{signal}}$ is the weight 
which implements the sideband background subtraction and $\Pepsilon^i=\Pepsilon(\mkpii,\cosks^i,\costhetaPsiGeni, \deltaphii)$ is the 
efficiency correction, obtained as described in \autoref{sec:selection}.

 \begin{figure}[t]
 \centering
 \begin{subfigure}[b]{.33\linewidth}
 \includegraphics[width=\linewidth]%
 {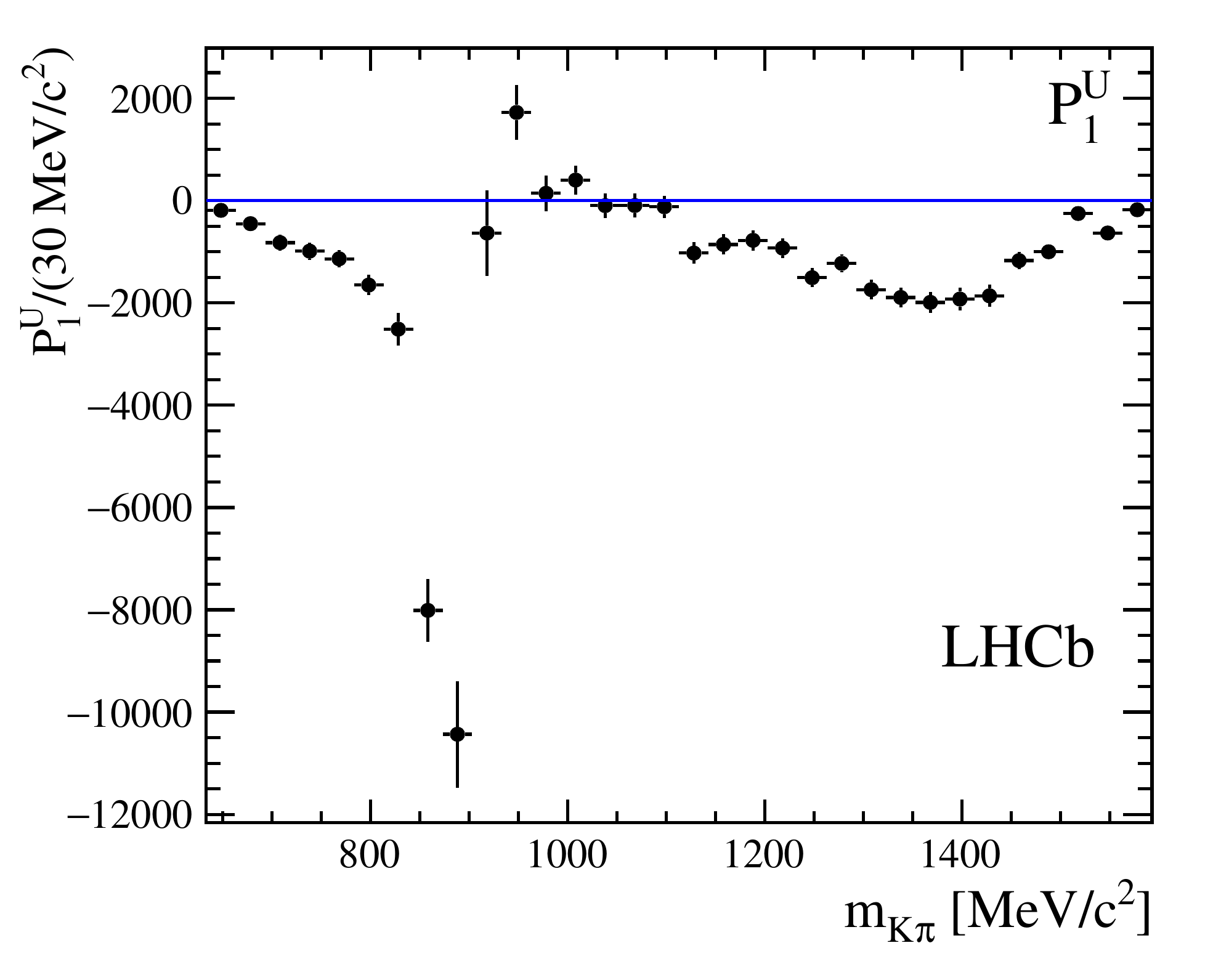}
 \end{subfigure}%
 \begin{subfigure}[b]{.33\linewidth}
 \includegraphics[width=\linewidth]%
 {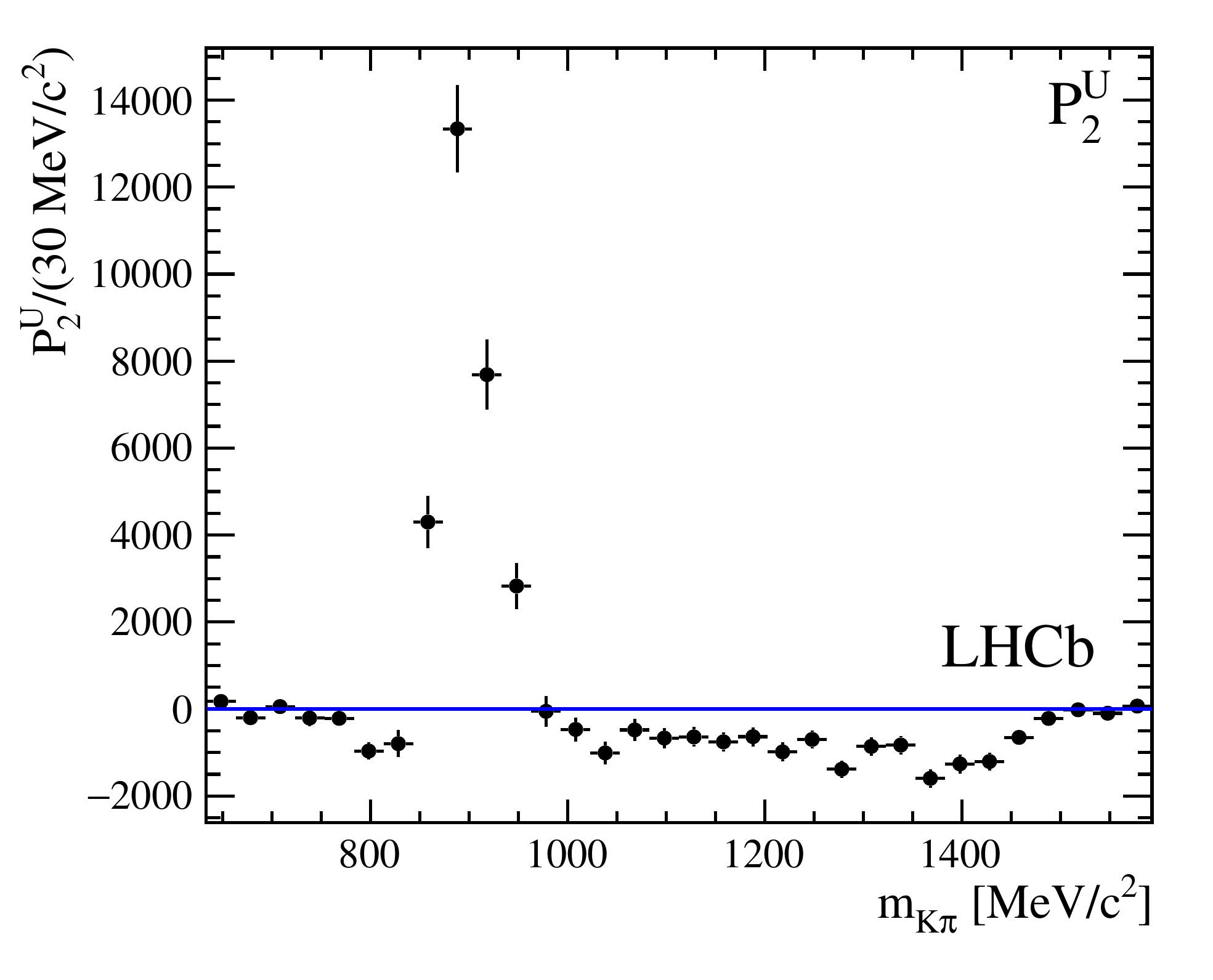}
 \end{subfigure}
 \begin{subfigure}[b]{.33\linewidth}
 \includegraphics[width=\linewidth]%
 {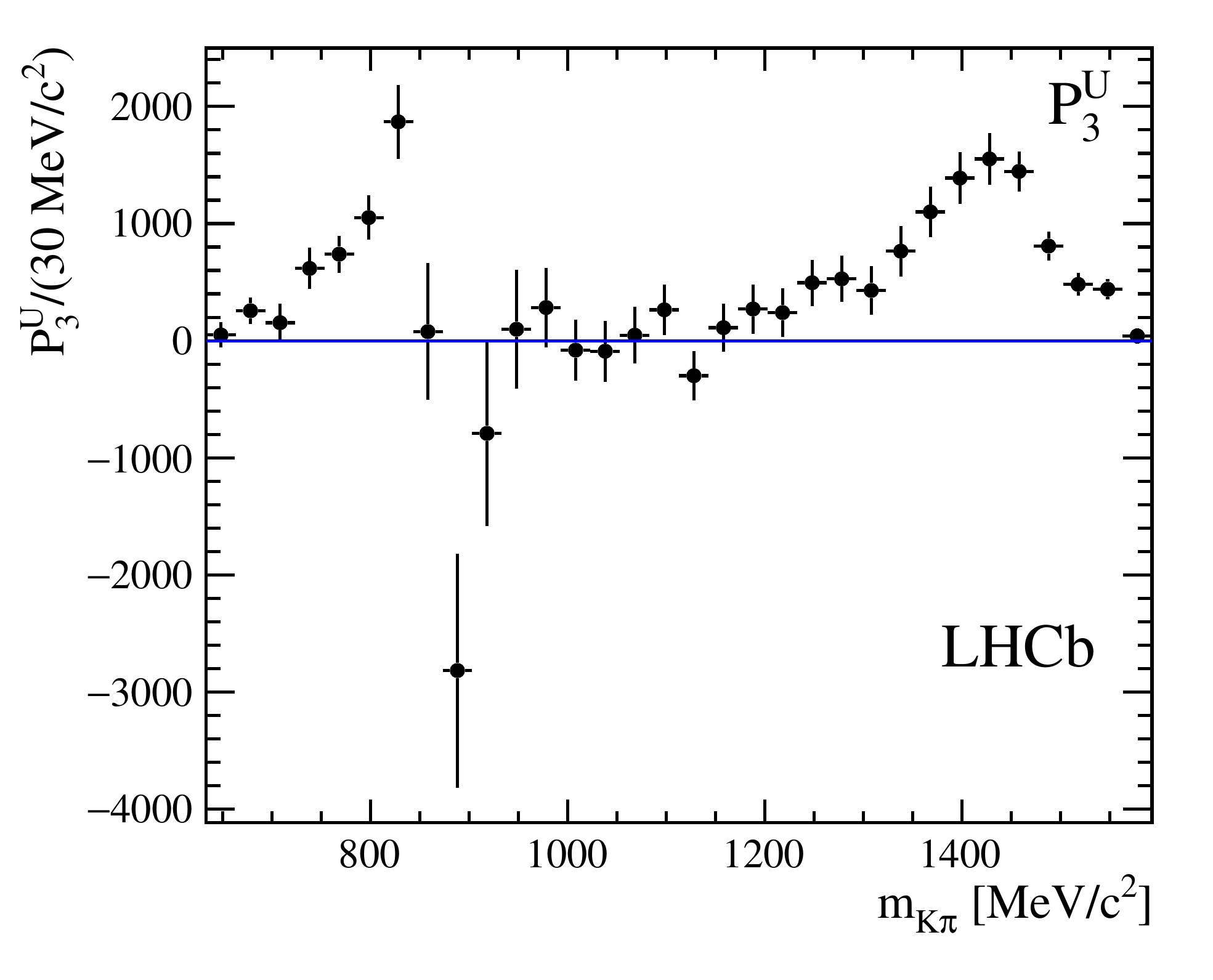}
 \end{subfigure}
 \\
 \begin{subfigure}[b]{.33\linewidth}
 \includegraphics[width=\linewidth]%
 {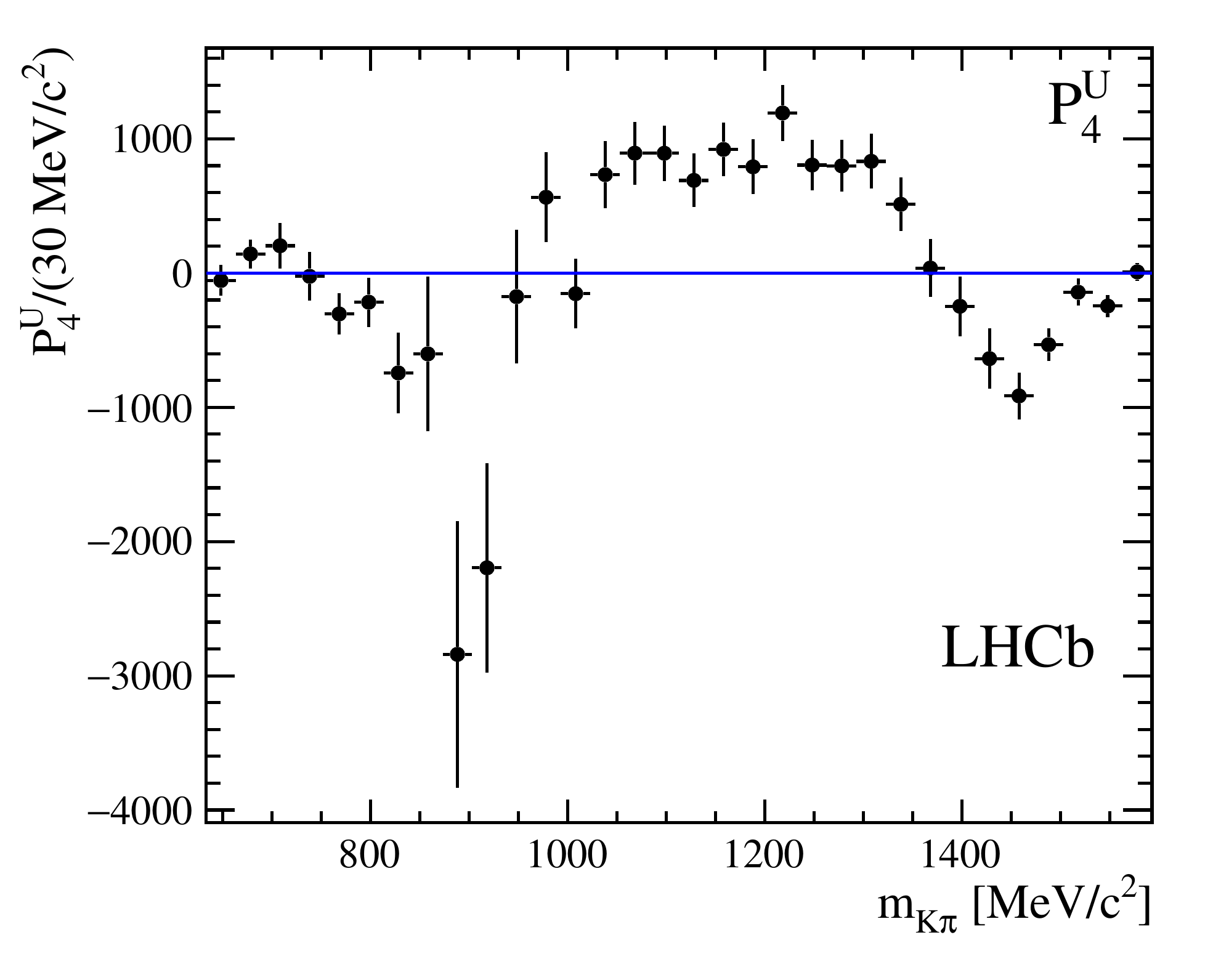}
 \end{subfigure}%
 \begin{subfigure}[b]{.33\linewidth}
 \includegraphics[width=\linewidth]%
 {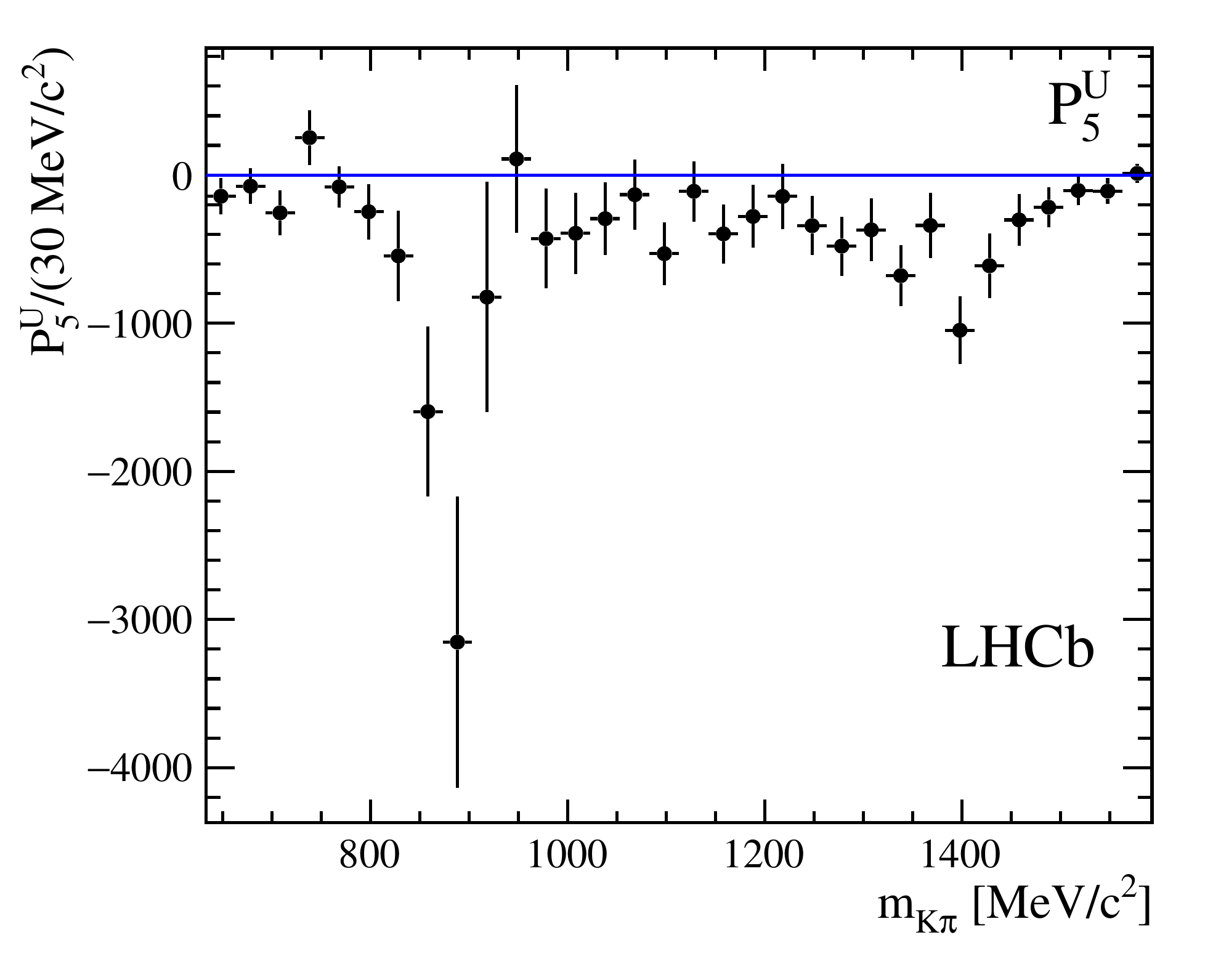}
 \end{subfigure}%
 \begin{subfigure}[b]{.33\linewidth}
 \includegraphics[width=\linewidth]%
 {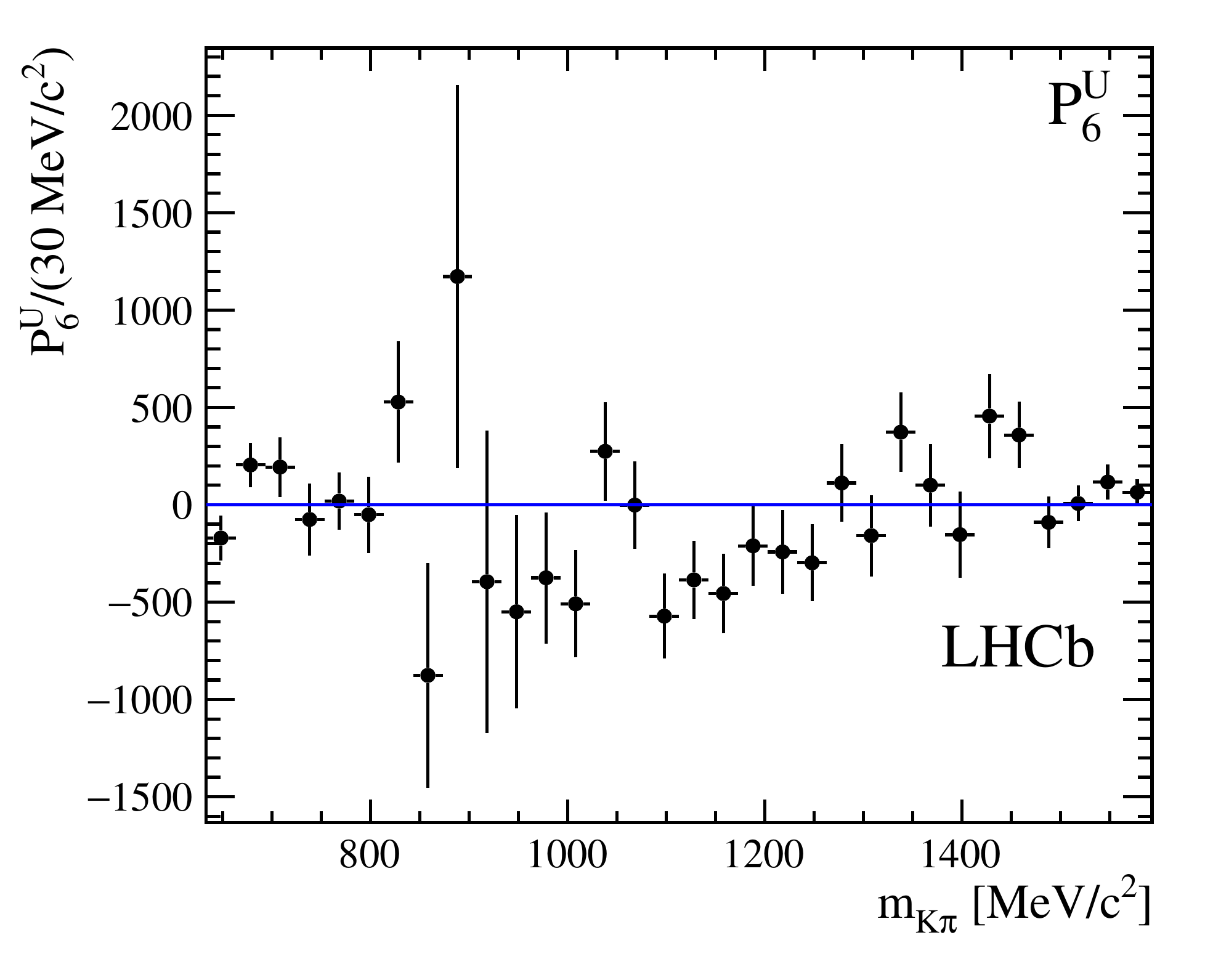}
 \end{subfigure}
 \caption{\small Dependence on \mkpi of the first six \Kpi moments of the \BzToKPiPsi decay mode as determined from data.}
 \label{fig:moments_B2KPiPsi}
 \end{figure}

The dependence of the first six moments on \mkpi is shown in \autoref{fig:moments_B2KPiPsi}.
Together with moment \pjua{0}, represented in the left plot of \autoref{fig:kstars_monodim}, moments \pjua{2} and \pjua{4} show the S, P and D wave amplitudes
in the mass regions of the  \Kstuno, \KstMQ, \Kstzero and \Kstdue resonances.  
The behavior of the moment \pjua{6}, generated by an F wave,
shows that any contribution from \Kstthree is small.
A resonant $\psipp$ state would, in general, contribute to all \Kpi moments.
\par A detailed discussion of these moments, together with the 
expressions relating  moments to the amplitudes, can be found in Ref. 
\cite{Aubert:2008aa} and references therein.


\ifthenelse{\boolean{prl}}{ }{ \tempclearpage }
\section{Analysis of the $\mpsipp$ spectrum}
\label{sec:MIReflectionOnMpsipi}
\begin{figure}[t]
\centering
\begin{subfigure}[b]{.33\linewidth}
\includegraphics[width=\linewidth]%
{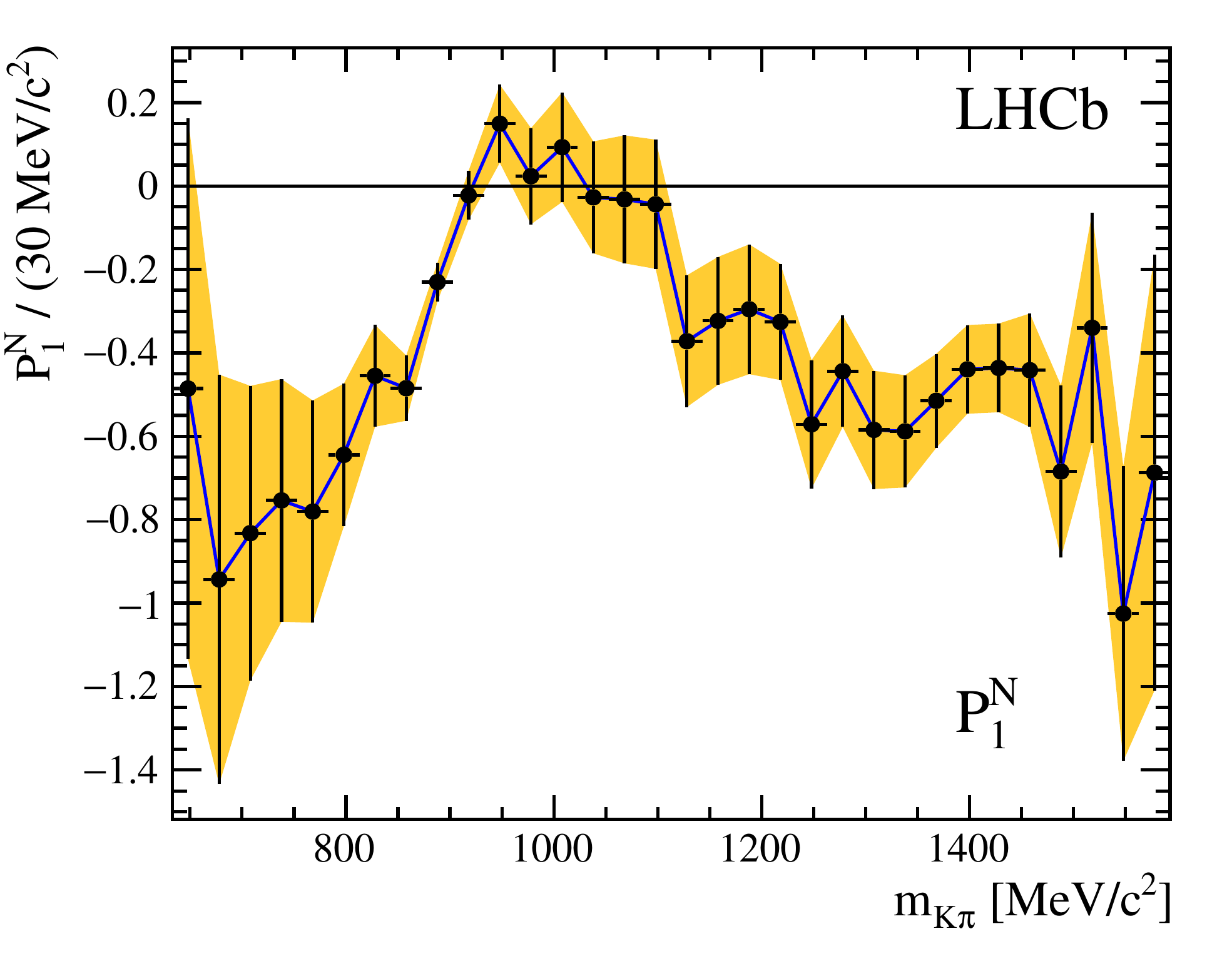}
\end{subfigure}%
\begin{subfigure}[b]{.33\linewidth}
\includegraphics[width=\linewidth]%
{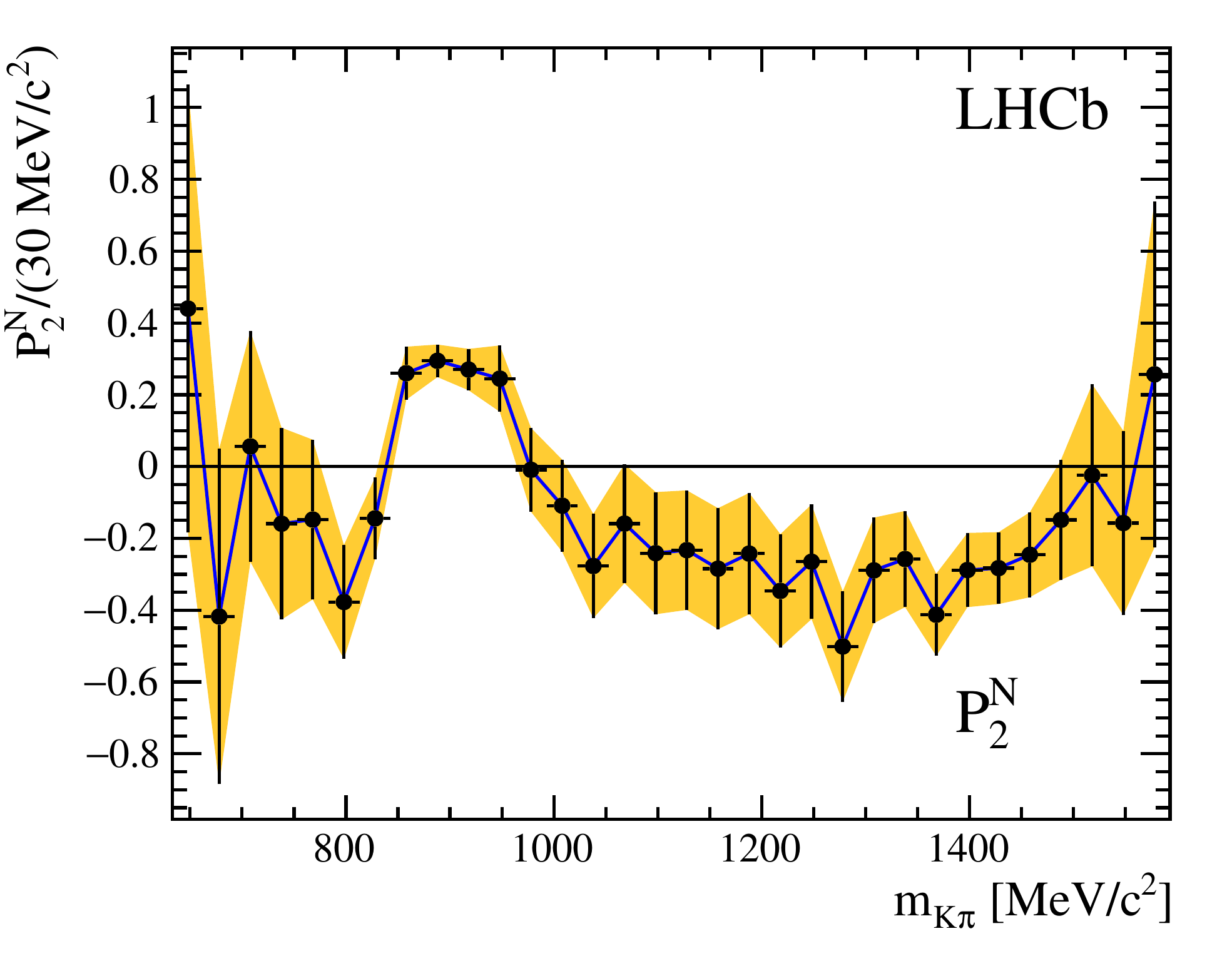}
\end{subfigure}%
\begin{subfigure}[b]{.33\linewidth}
\includegraphics[width=\linewidth]%
{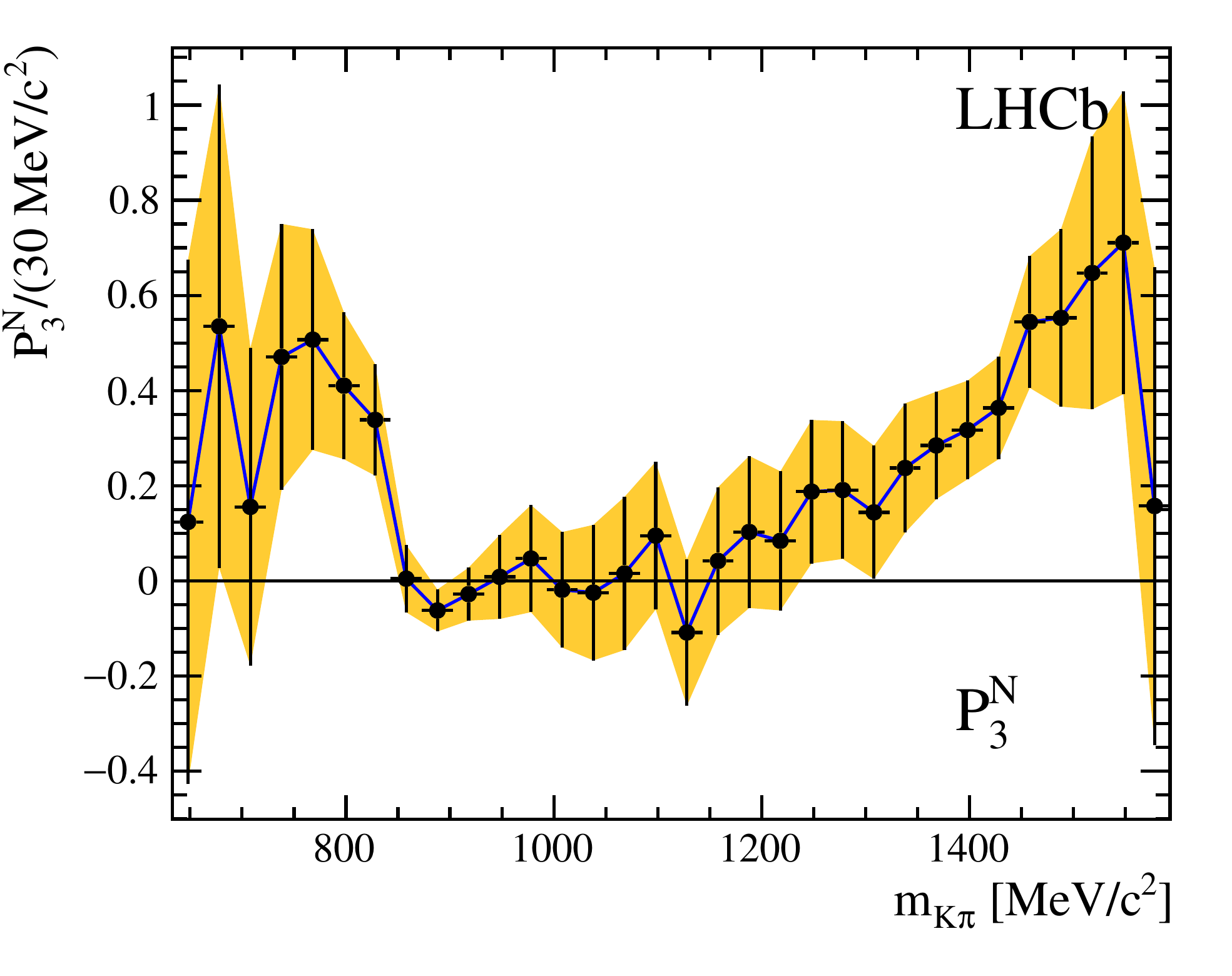}
\end{subfigure}
\\
\begin{subfigure}[b]{.33\linewidth}
\includegraphics[width=\linewidth]%
{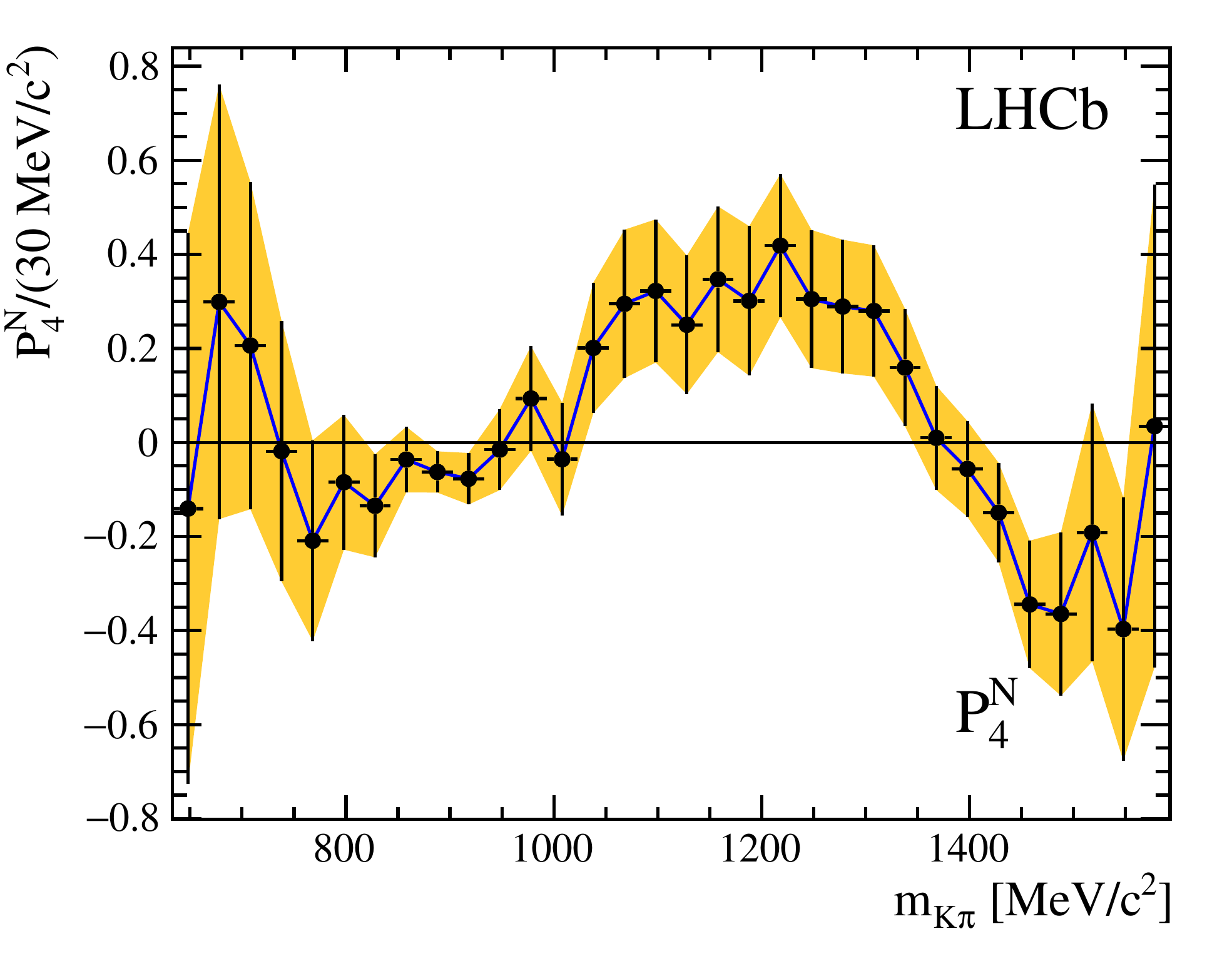}
\end{subfigure}%
\begin{subfigure}[b]{.33\linewidth}
\includegraphics[width=\linewidth]%
{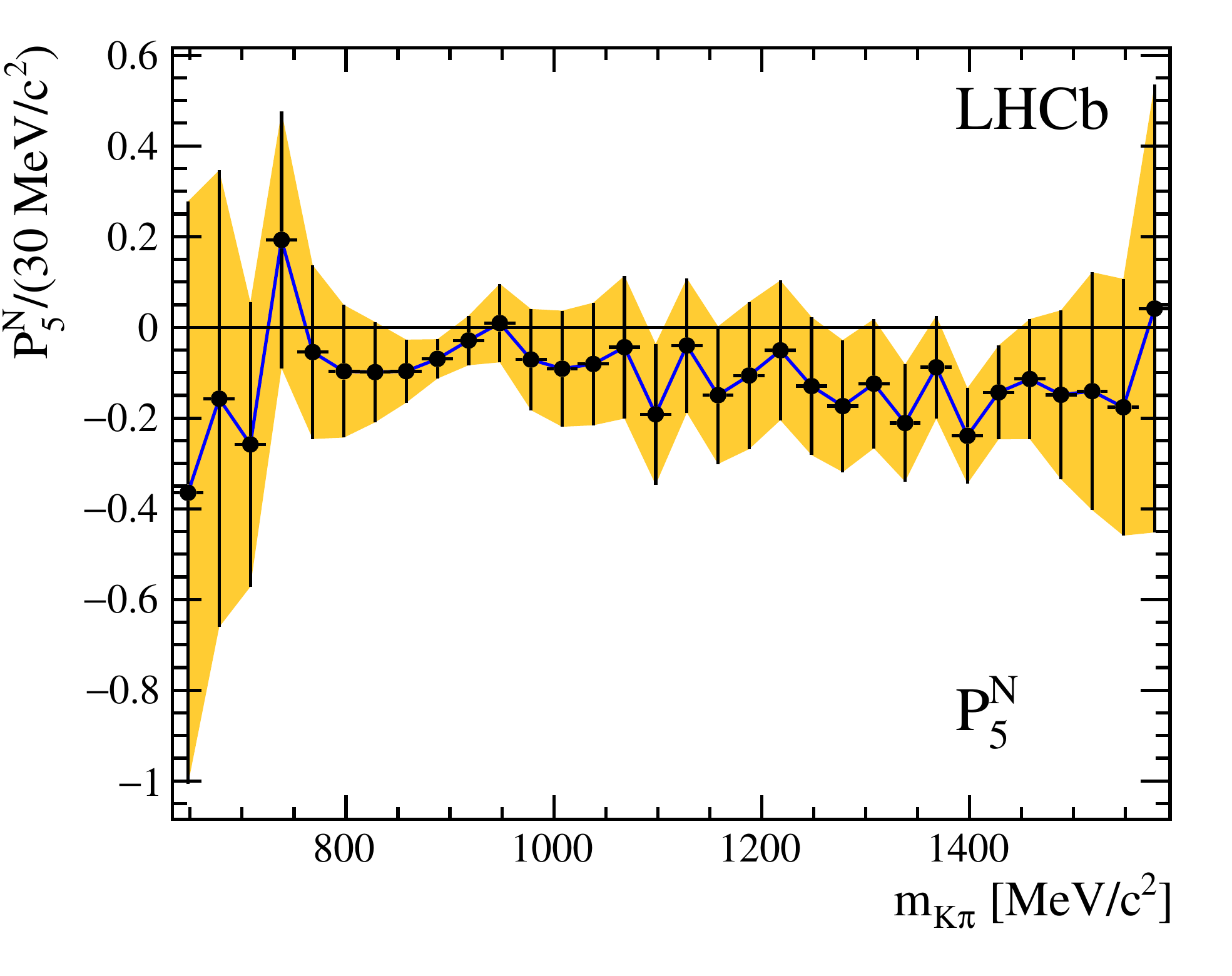}
\end{subfigure}%
\begin{subfigure}[b]{.33\linewidth}
\includegraphics[width=\linewidth]%
{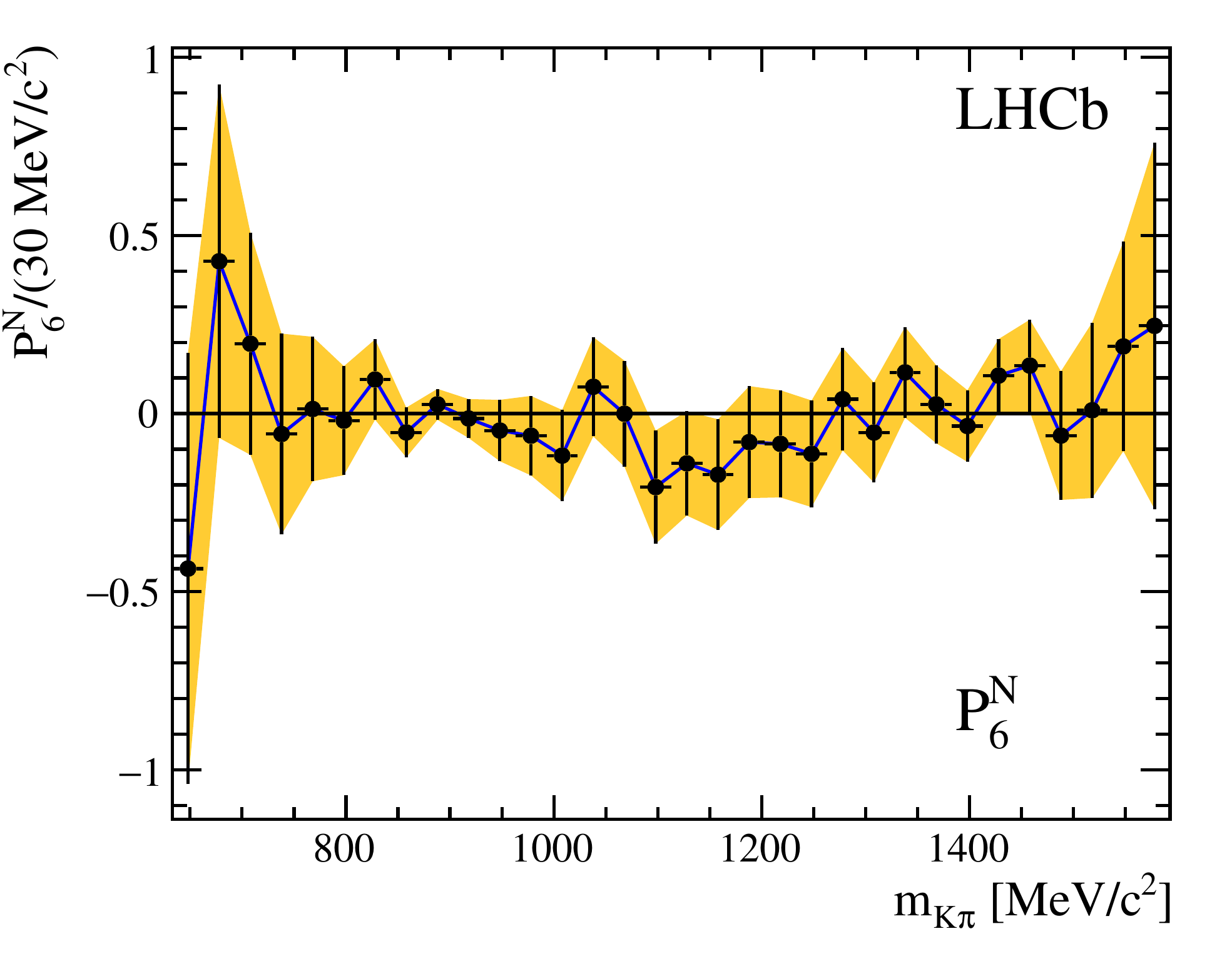}
\end{subfigure}
\caption{\small First six normalized \kpi moments of the \BzToKPiPsi decay mode as a function of \mkpi. The shaded (yellow) bands indicate the $\pm 1\sigma$ variations of the moments.}
\label{fig:norm_moments_B2KPiPsi}
\end{figure}

\begin{figure}[t]
\centering
\begin{subfigure}[b]{.5\linewidth}
\includegraphics[width=\linewidth]%
{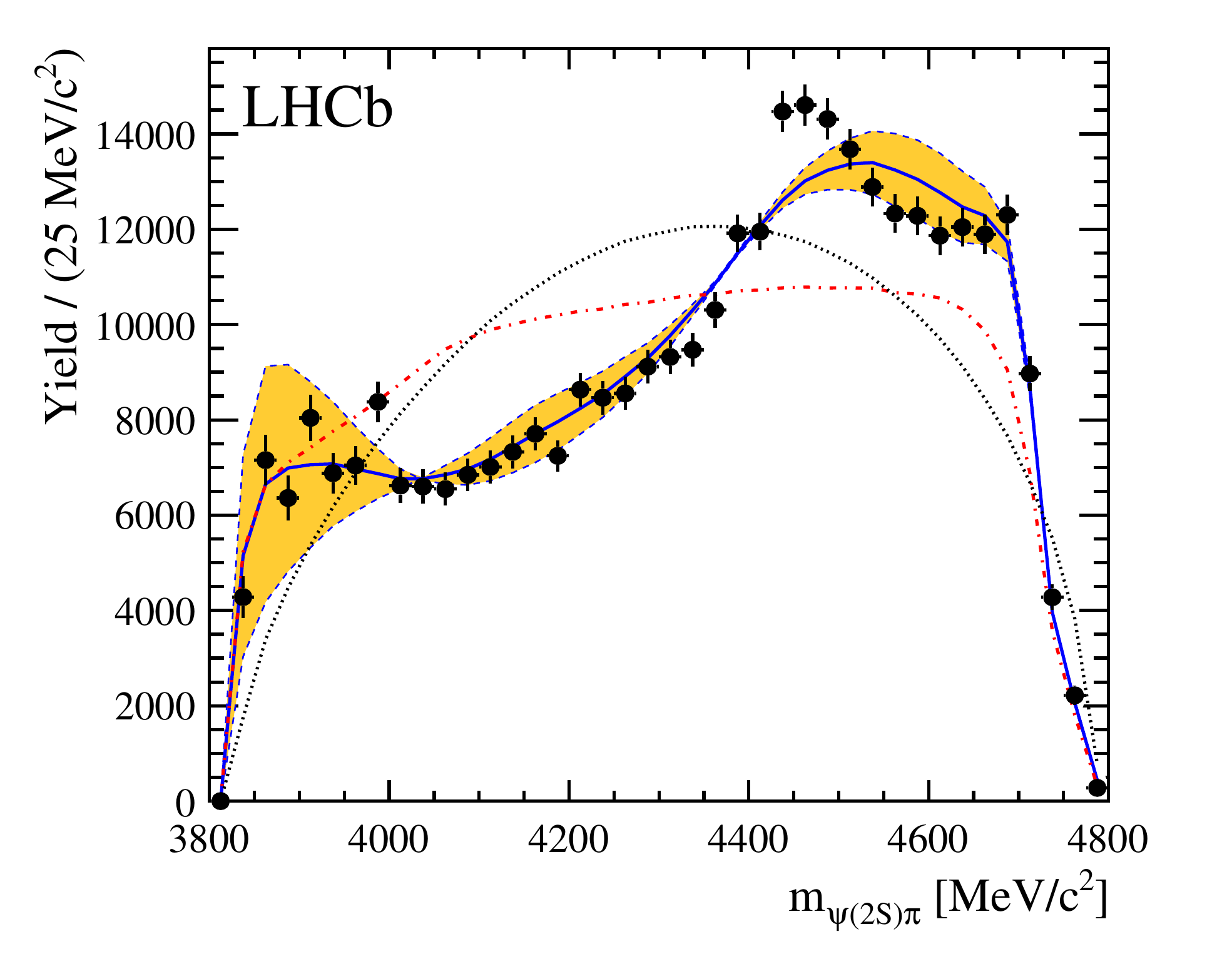}
\end{subfigure}%
\begin{subfigure}[b]{.5\linewidth}
\includegraphics[width=\linewidth]%
{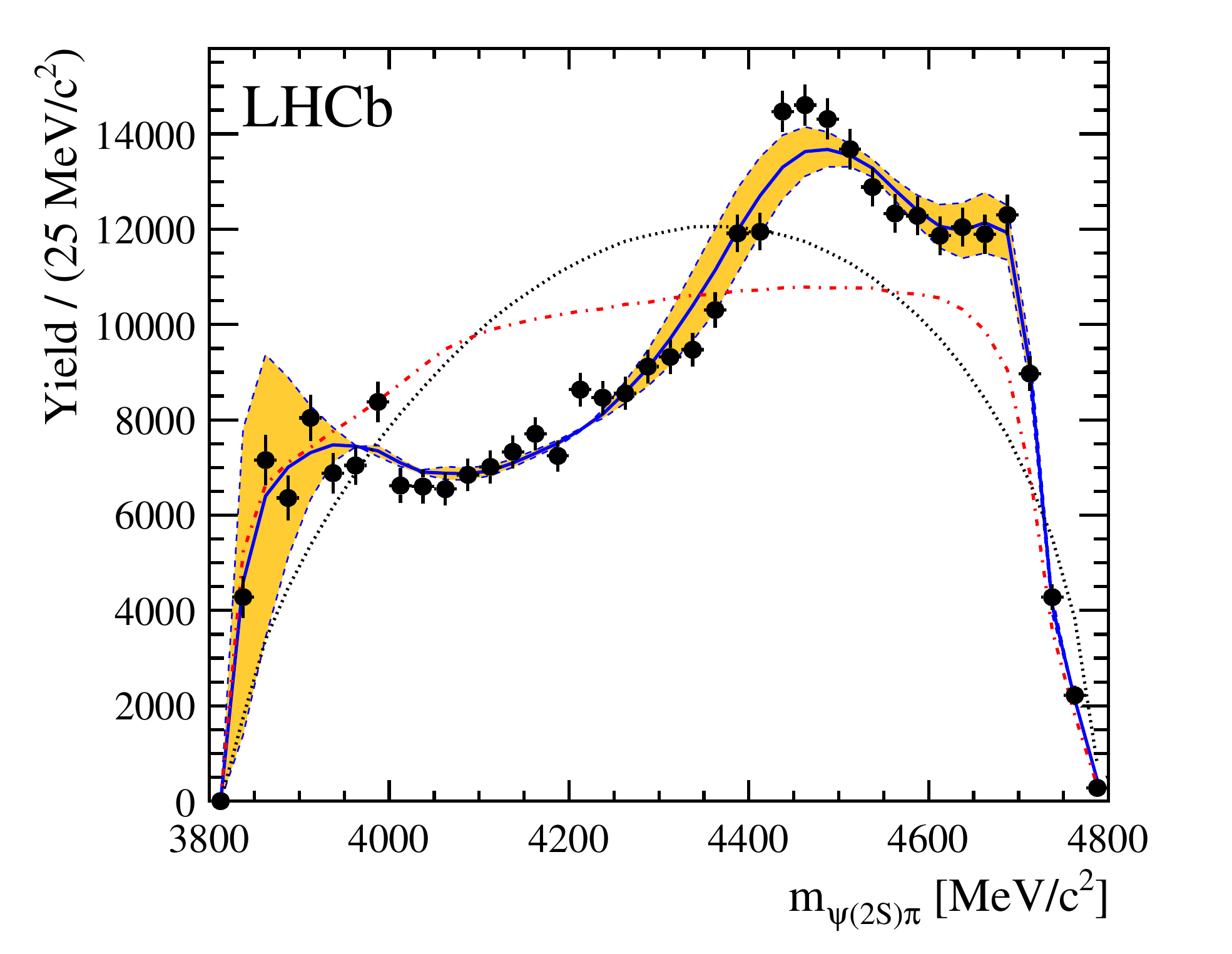}
\end{subfigure}%
\caption{ \small Background subtracted and efficiency corrected spectrum of $\mpsipp$. Black points represent data. Superimposed are the distributions of the Monte Carlo simulation: the dotted (black) line corresponds to the pure phase-space case; in the dash-dotted (red) line the \mkpi spectrum is weighted to reproduce the experimental distribution; in the continuous (blue) line the angular structure of the \kpi system is incorporated using Legendre polynomials up to (left) $\lmax=4$ and (right) $\lmax=6$. 
The shaded (yellow) bands are related to the uncertainty on normalized moments, which is due to the statistical uncertainty that comes from the data. Therefore the two uncertainties should not be combined when comparing data and Monte Carlo predictions. See text for further details.}
\label{fig:PiPsi}
\end{figure}

\begin{figure}[t]
\centering
\includegraphics[width=\linewidth]%
{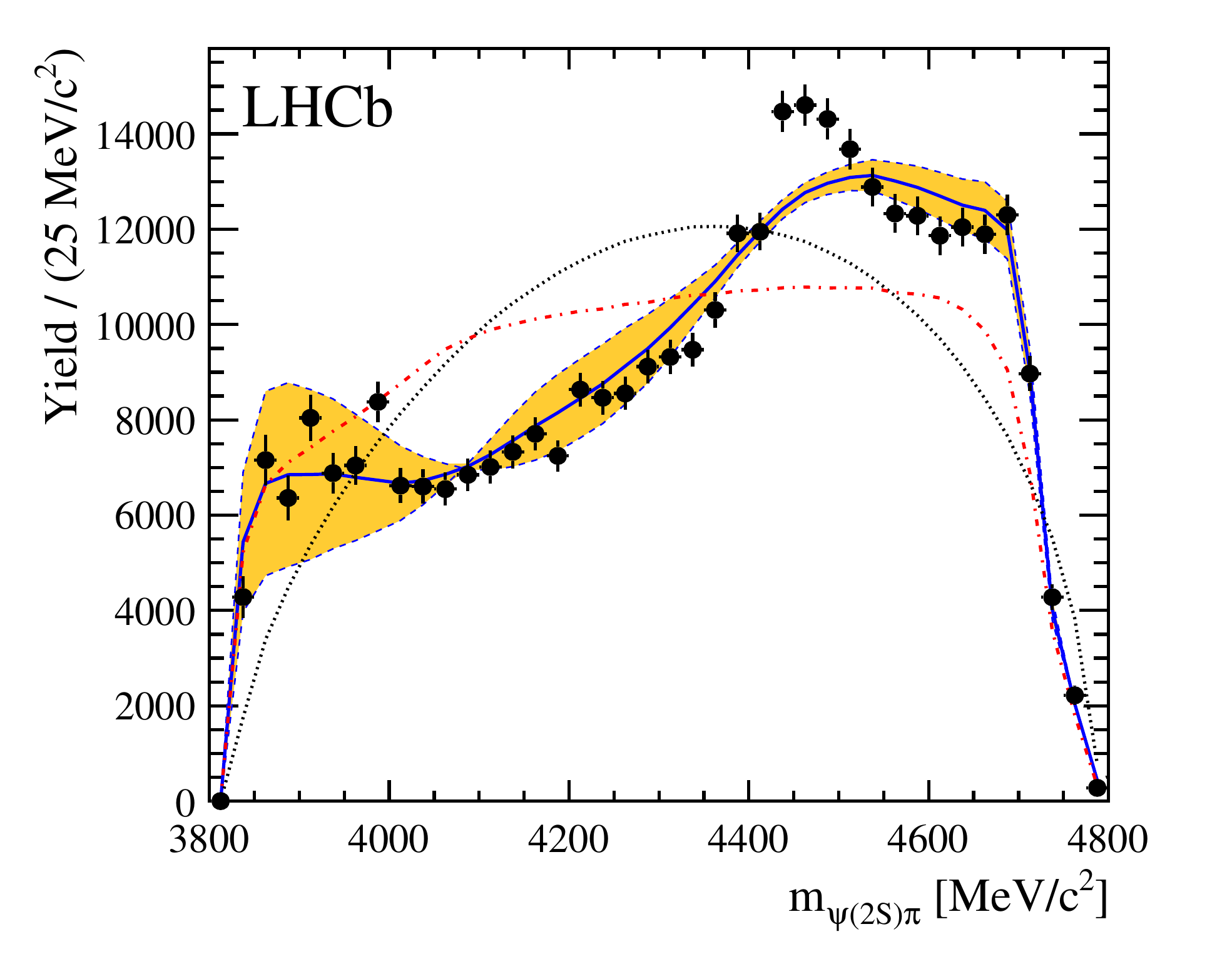}
\caption{ \small The experimental spectrum of $\mpsipp$ is shown by the black points. Superimposed are the distributions of the Monte Carlo simulation: the dotted(black) line corresponds to the pure phase-space case; in the dash-dotted (red) line the \mkpi spectrum is weighted to reproduce the experimental distribution; in the continuous (blue) line the angular structure of the \kpi system is incorporated using Legendre polynomials with index 
$\lmax$ variable according to \mkpi as described in  ~\autoref{eq:slices}, reaching up to $\lmax=4$. The shaded (yellow) bands
are related to the uncertainty on normalized moments, which is due 
to the statistical uncertainty that comes from the data. Therefore the two uncertainties should not be combined when comparing data and Monte Carlo predictions. See text for further details.}
\label{fig:psipi_Lmaxvar}
\end{figure}

The reflection of the mass and angular structure 
of the $K\pi$ system into the $\psipp$ invariant mass  
spectrum is investigated to establish whether it is sufficient
to explain the data distribution.
This is achieved by comparing the experimental
$\mpsipp$ spectrum to that of a simulated sample which accounts for the measured mass spectrum and the angular distribution of the \Kpi system by means of appropriate weights.
The comparison is performed in three configurations of the \Kpi spin contributions.
The simplest configuration corresponds to including the contributions of S, P and D waves, which account for all resonances with mass below the kinematic limit
and the \KstMS meson, just above it (see \autoref{tab:kstars}). In the second configuration the $\Kstthree$ meson  is also allowed to contribute. 
This represents a rather unlikely assumption since it implies a sizeable presence of spin-3 resonances at 
low \mkpi. This configuration can be considered as an extreme case that provides a valuable test for the robustness of the method.
In the third configuration a more realistic choice is made by limiting  
the maximum spin as a function of \mkpi.

For each of the three configurations, 50 million simulated events are generated according to the $\BzToKPiPsi$ phase-space decay. 
The simulation does not include detector effects because it will be compared to efficiency-corrected data. The simulated \mkpi distribution is forced to reproduce the \Kpi spectrum in data (left plot of \autoref{fig:kstars_monodim})  by attributing to each event 
a weight proportional to the ratio between the real and simulated \mkpi spectra in the appropriate bin.
Finally, the  angular structure of the \Kpi system is modified in the simulated sample 
by applying an additional weight to each event computed as

\begin{equation}
\label{eq:weight}
    w^i = 1 + \sum_{j=1}^{\lmax}\pjn P_j(\cosks^i),
\end{equation}

\noindent where $\pjn = 2\pju/N_{\text{corr}}$ are the normalized moments, derived from the moments \pju of \autoref{eq:unn_moments}, and $N_{\rm corr}$ is the background-subtracted and efficiency-corrected yield of the \mkpi bin where the event lies.
The behavior of the first six normalized moments is shown in \autoref{fig:norm_moments_B2KPiPsi}. The value and the uncertainty of these moments, at a given \mkpi value, are estimated by linearly interpolating adjacent points and their $\pm 1\sigma$ values, respectively, as shown by the shaded (yellow) bands in the figures. 
\par The experimental distribution of the $\psipp$ system invariant mass, 
$\mpsipp$, is shown by the black points in the left plot of \autoref{fig:PiPsi}.\footnote{This plot uses an improved parametrization of the \Bz mass spectrum with respect to Fig.~1 in Ref. \cite{LHCb-PAPER-2014-014}.}
The dotted (black) line represents the pure phase-space simulation,
the dash-dotted (red) line shows the effect of the \mkpi modulation, while in the continuous (blue) line the angular structure
of the \Kpi system has been taken into account by allowing S, P and D waves 
to contribute, which corresponds to setting $\lmax=4$ in \autoref{eq:weight}.
The effect of the angular structure of the \kpi system accounts for most of the features seen
in the $\mpsipp$ spectrum except for the peak around $4430\mevcc$.
The dashed (yellow) band in the figure is derived from the $\pm 1\sigma$ values of the normalized moments. The  borders of the band are calculated by attributing to each simulated event the weight in \autoref{eq:weight} assuming the values of $+1\sigma$ or $-1\sigma$, simultaneously for all the contributing normalized moments. 
Due to the negative contributions of the moments, the borders may cross the central continuous (blue) line. The band should not be considered as an uncertainty in the simulation but only as an indicative measure of the limited data sample used to compute moments. Since the band and the error bars on the black points are related to the same statistical uncertainty on the data, they should not be combined when estimating the statistical significance of deviations of the data from the prediction.

\begin{figure}[t]
\centering
\begin{subfigure}[b]{.45\linewidth}
\includegraphics[width=\linewidth]%
{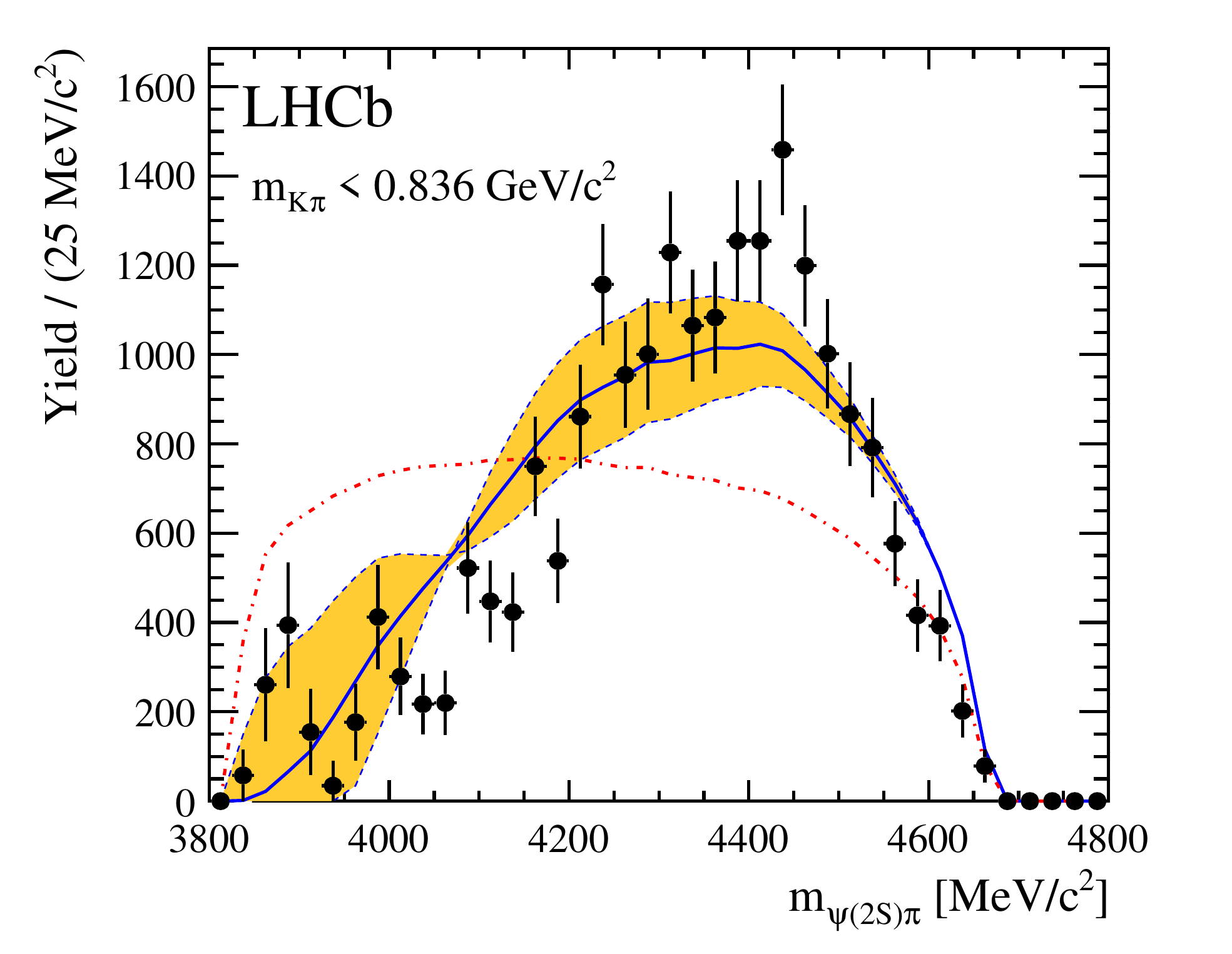}
\end{subfigure}%
\begin{subfigure}[b]{.45\linewidth}
\includegraphics[width=\linewidth]%
{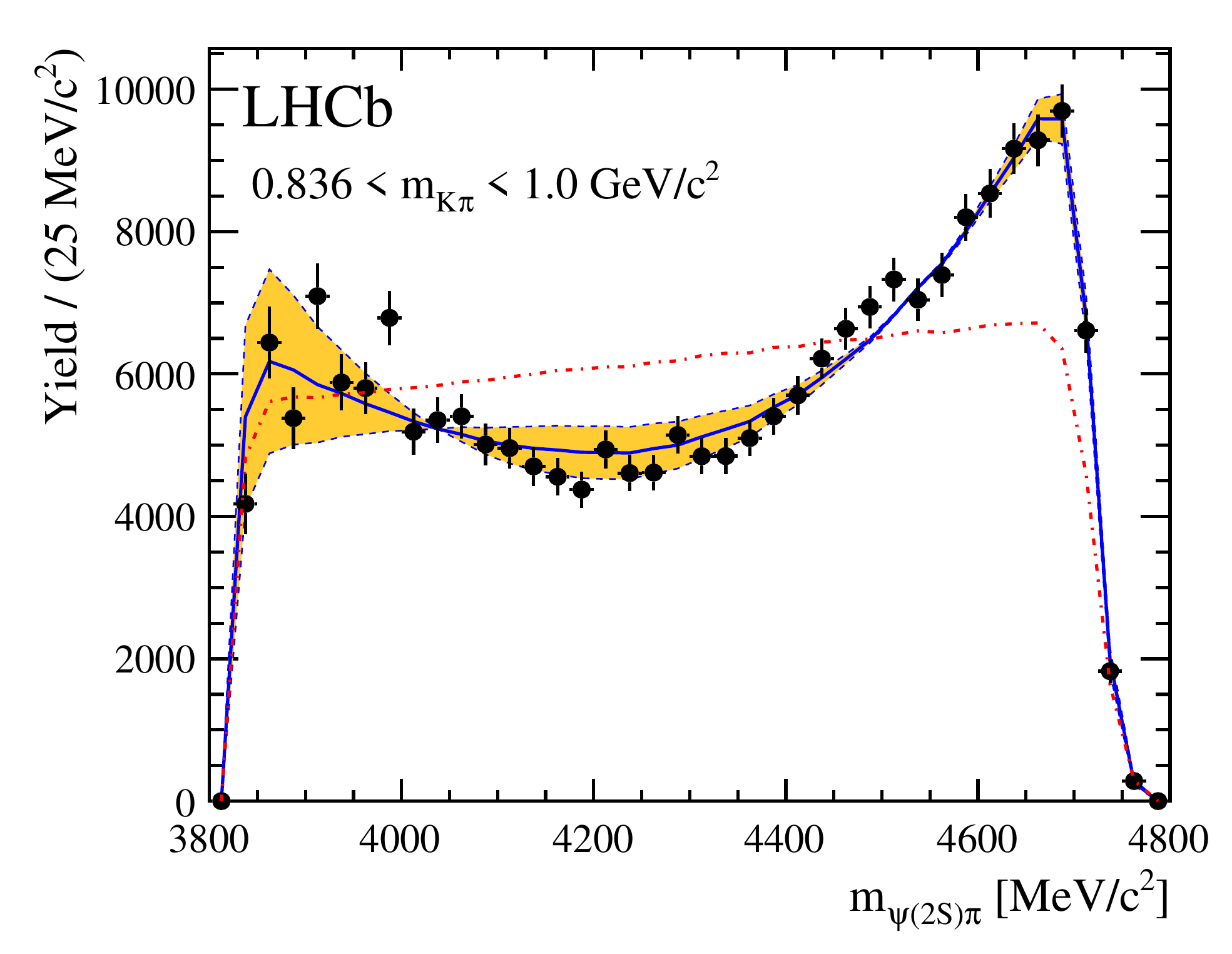}
\end{subfigure}%
\\
\begin{subfigure}[b]{.45\linewidth}
\includegraphics[width=\linewidth]%
{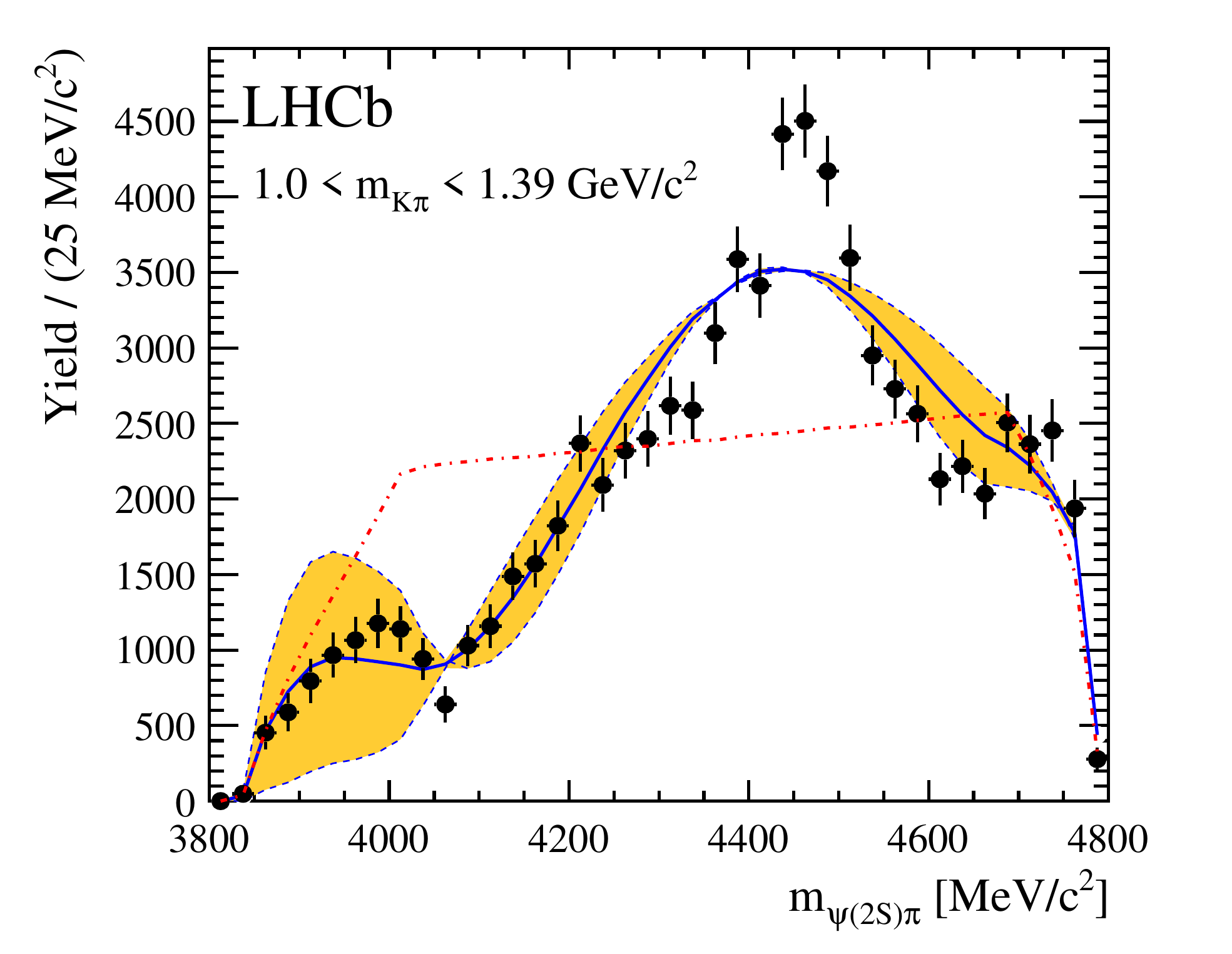}
\end{subfigure}
\begin{subfigure}[b]{.45\linewidth}
\includegraphics[width=\linewidth]%
{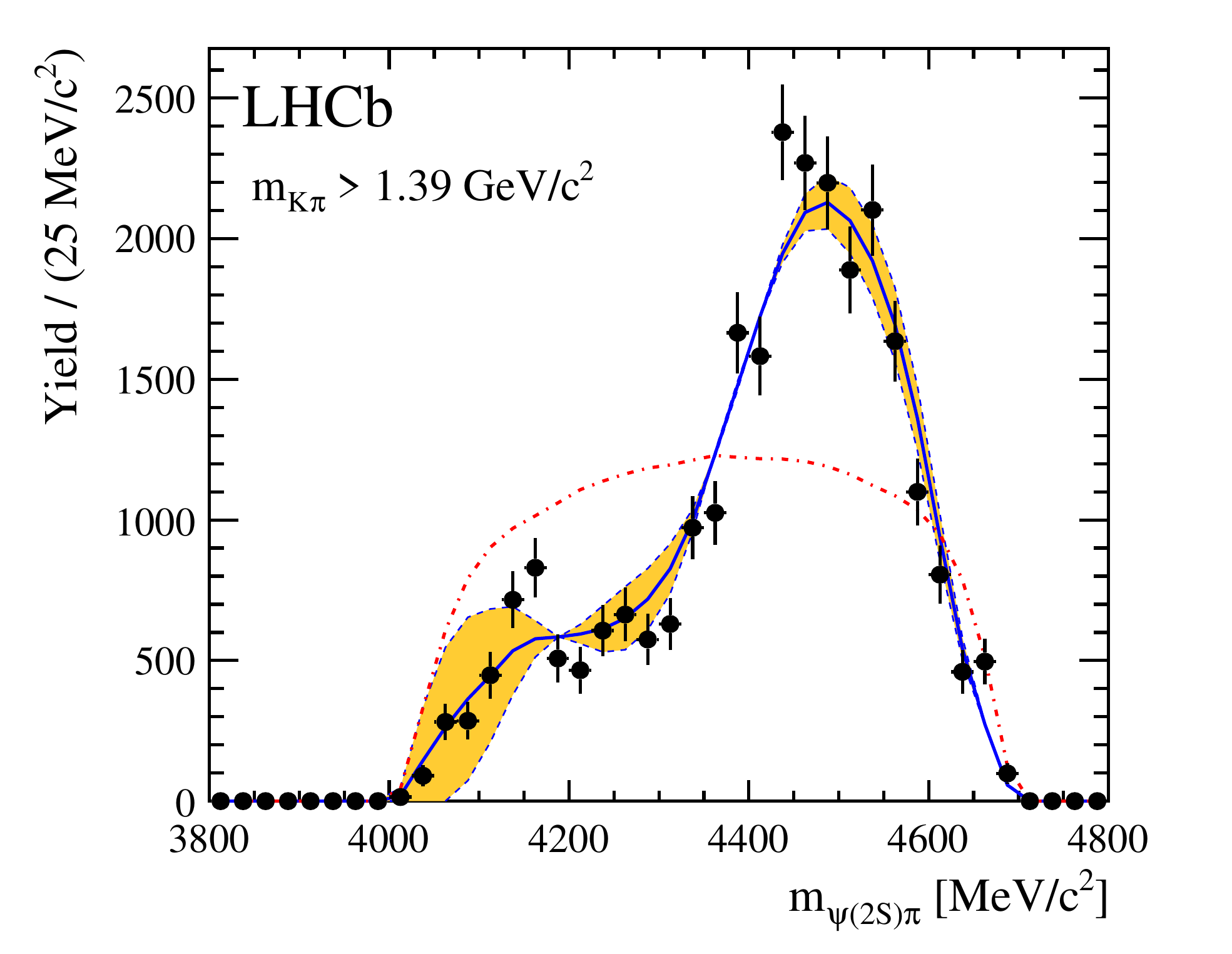}
\end{subfigure}%
\captionsetup{skip=1pt}
\caption{\small Black points represent the experimental distributions of $\mpsipp$ for the indicated \mkpi intervals.
The dash-dotted (red) line is obtained by modifying the \mkpi spectrum of the phase-space simulation according to the \mkpi experimental spectrum. In the continuous (blue) line the angular structure of the \kpi system is incorporated using Legendre polynomials with variable index $\lmax$ according to \autoref{eq:slices}. The shaded (yellow) bands
are related to the uncertainty on normalized moments, which is due 
to the statistical uncertainty that comes from the data.
Therefore the two uncertainties should not be combined when comparing data and Monte Carlo predictions. See text for further details. }
\label{fig:psipi_lmaxvar-slices}
\end{figure}

When spin-3 \kpi states are included, by setting $\lmax=6$, 
the  predicted $\mpsipp$ spectrum is  modified as shown on the right plot of \autoref{fig:PiPsi}.
Even though the $\lmax=6$ solution apparently provides a better description of the data, it is shown in the following that it is largely incompatible with the data.
\par In \autoref{fig:psipi_Lmaxvar}
the maximum Legendre polynomial order is limited as  a function of \mkpi, according to
 \begin{equation}
\label{eq:slices}
   \lmax = \left\{
     \begin{array}{lc}
       2 & \mkpi<836\mevcc\\
       3 & 836\mevcc<\mkpi<1000\mevcc\\
       4 & \mkpi >1000\mevcc.\\
     \end{array}
   \right.
\end{equation} 
\noindent Figure \ref{fig:psipi_Lmaxvar}
demonstrates that with this better-motivated $\lmax$ assignment, the simulation cannot reproduce adequately the $\mpsipp$ distribution.
\par The disagreement is more evident when looking at the same spectra in different intervals of
\mkpi, as shown in \autoref{fig:psipi_lmaxvar-slices}. 
Here the candidates are subdivided according to the \mkpi intervals defined in \autoref{eq:slices}.
The last interval is further split into $1000\mevcc<\mkpi<1390\mevcc$ and $\mkpi>1390\mevcc$. 
Except for the mass region around $4430\mevcc$, all slices exhibit good agreement between the data and the simulation. The peaking structure is particularly evident in the region $1000\mevcc<\mkpi<1390\mevcc$, between the
\Kstuno and the resonances above $1400\mevcc$.

\vskip 1cm


\ifthenelse{\boolean{prl}}{ }{ \tempclearpage }
\section{Statistical significance of the result}
\label{sec:MIHypothesistest}
\begin{figure}[t]
\centering
\includegraphics[width=.6\linewidth]%
{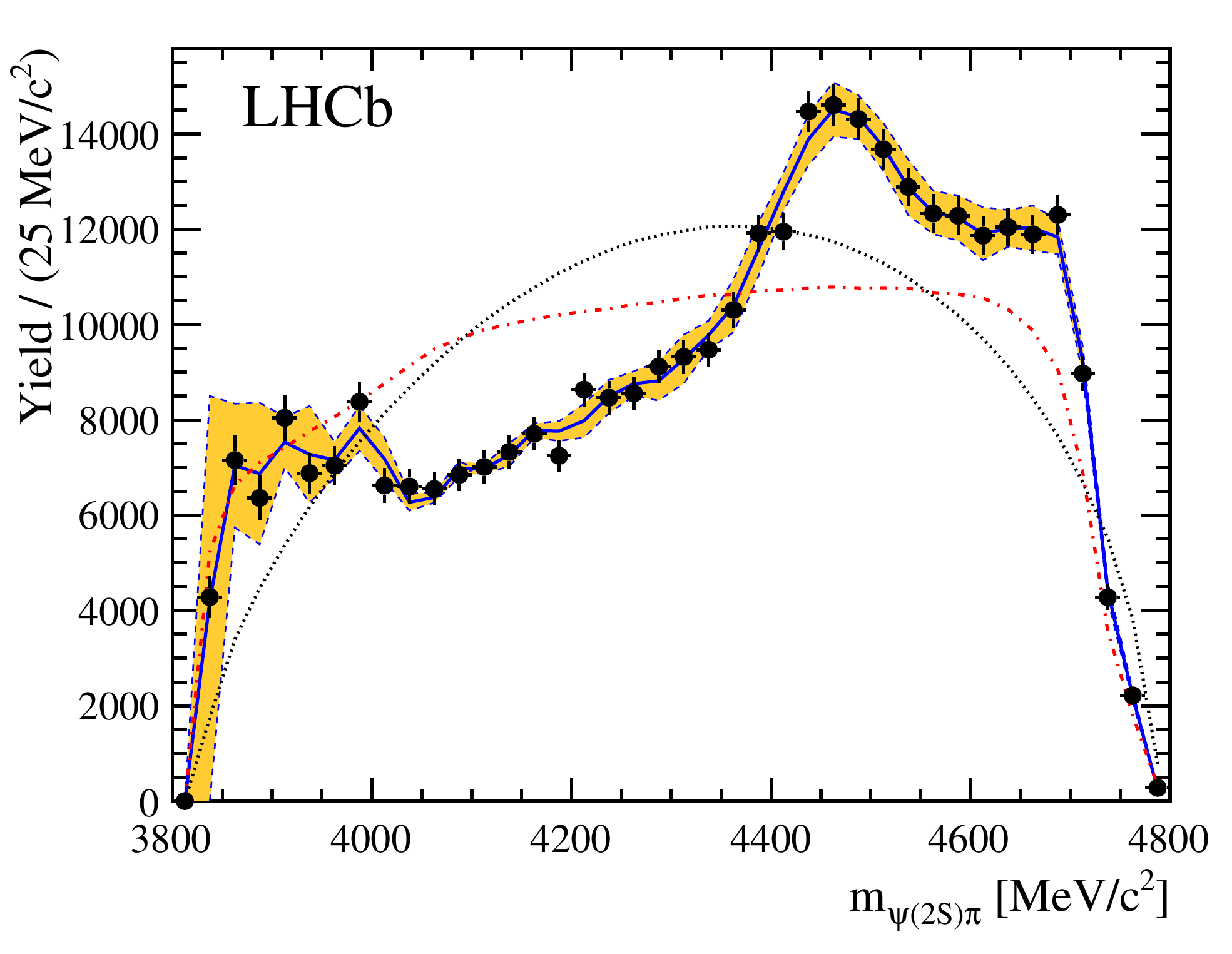}
\caption{ \small The experimental spectrum of $\mpsipp$ is shown by the black points. Superimposed are the distributions of the Monte Carlo simulation: the dotted (black) line corresponds to the pure phase space case; in the dash-dotted (red) line the \mkpi spectrum is weighted to reproduce the experimental distribution; in the continuous (blue) line the angular structure of the \kpi system is incorporated using Legendre polynomials up to 
$\lmax=30$ which implies a full description of the spectrum features even if it corresponds to an unphysical configuration of the \Kpi system. 
The shaded (yellow) bands are related to the uncertainty on normalized moments.}
\label{fig:lmax30}
\end{figure}

The fact that  the $\mpsipp$ spectrum cannot be explained  as a reflection of 
the angular structure of the \Kpi system has been illustrated qualitatively. In this section, the disagreement is quantified via a hypothesis-testing procedure using a likelihood-ratio estimator.
The compatibility between the expected $\mpsipp$ distribution,
accounting for the reflections of the \Kpi angular structure, and that
observed experimentally is tested for the three $\lmax$ assignments described in the previous section, with
three sets of about 1000 pseudoexperiments each.
For each pseudoexperiment, data and simulated samples involved in the analysis chain are reproduced as pseudosamples, generated at the same statistical level as in the real case.
The signal candidate pseudosamples are extracted from a $({\mkpi},\cosks,\costhetaPsiGen, \deltaphi)$  distribution obtained by an independent EvtGen \cite{Lange:2001uf} phase-space sample of $\BzToKPiPsi$ events. The distribution is generated, for each of the three $\lmax$ cases previously discussed, in order to reproduce the (\mkpi,$\cosks$) behavior. 
The background pseudosamples are simulated according to the $({\mkpi},\cosks,\costhetaPsiGen, \deltaphi)$  distribution of the candidates in 
the $\Bz$ invariant mass side-bands. 
Finally, to mimic the calculation of the efficiency correction factors, 
two additional samples are generated by extracting events from two distributions, in the same 4D space, obtained from the simulation with full detector effects, before and after the application of the analysis chain.  
The sum of the signal and background samples is then subject to background
subtraction and efficiency correction, exactly as for the real data, and moments are calculated.
In the pseudoexperiments, events are simulated with equal amounts of each $\psip$ polarisation state.
Effects related to $\psip$ polarization are only included in the pseudosample via their correlation with the \Kpi mass and $\cosks$ distributions, which are derived from data.   
It has been checked that this does not significantly influence the results although 
the validity of such approximate treatment of $\psip$ polarization is, in general, analysis dependent and not necessarily appropriate in other experimental situations.
\begin{figure}[t]
\centering
\begin{subfigure}[b]{.33\linewidth}
\includegraphics[width=\linewidth]%
{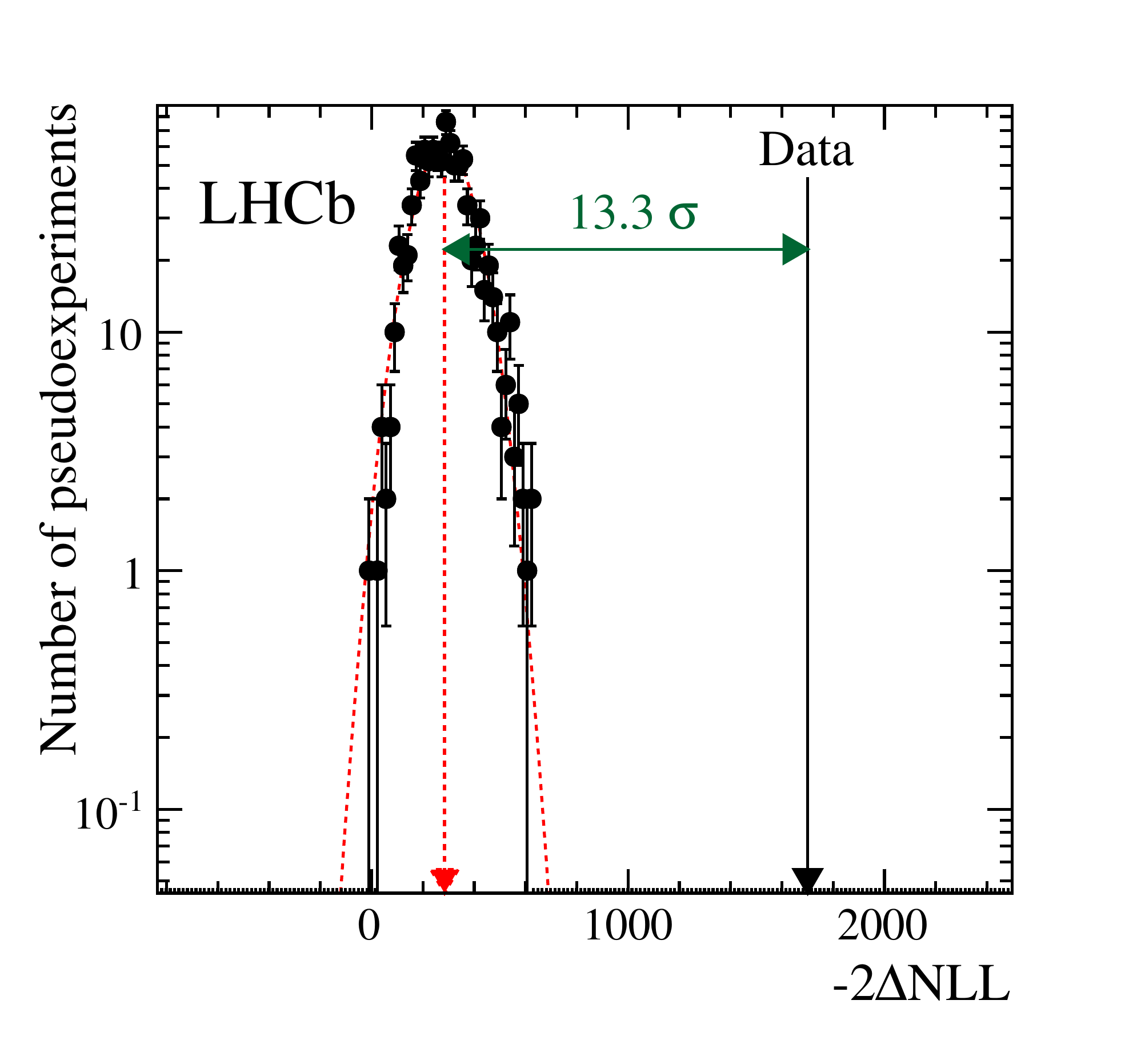}
\end{subfigure}%
\begin{subfigure}[b]{.33\linewidth}
\includegraphics[width=\linewidth]%
{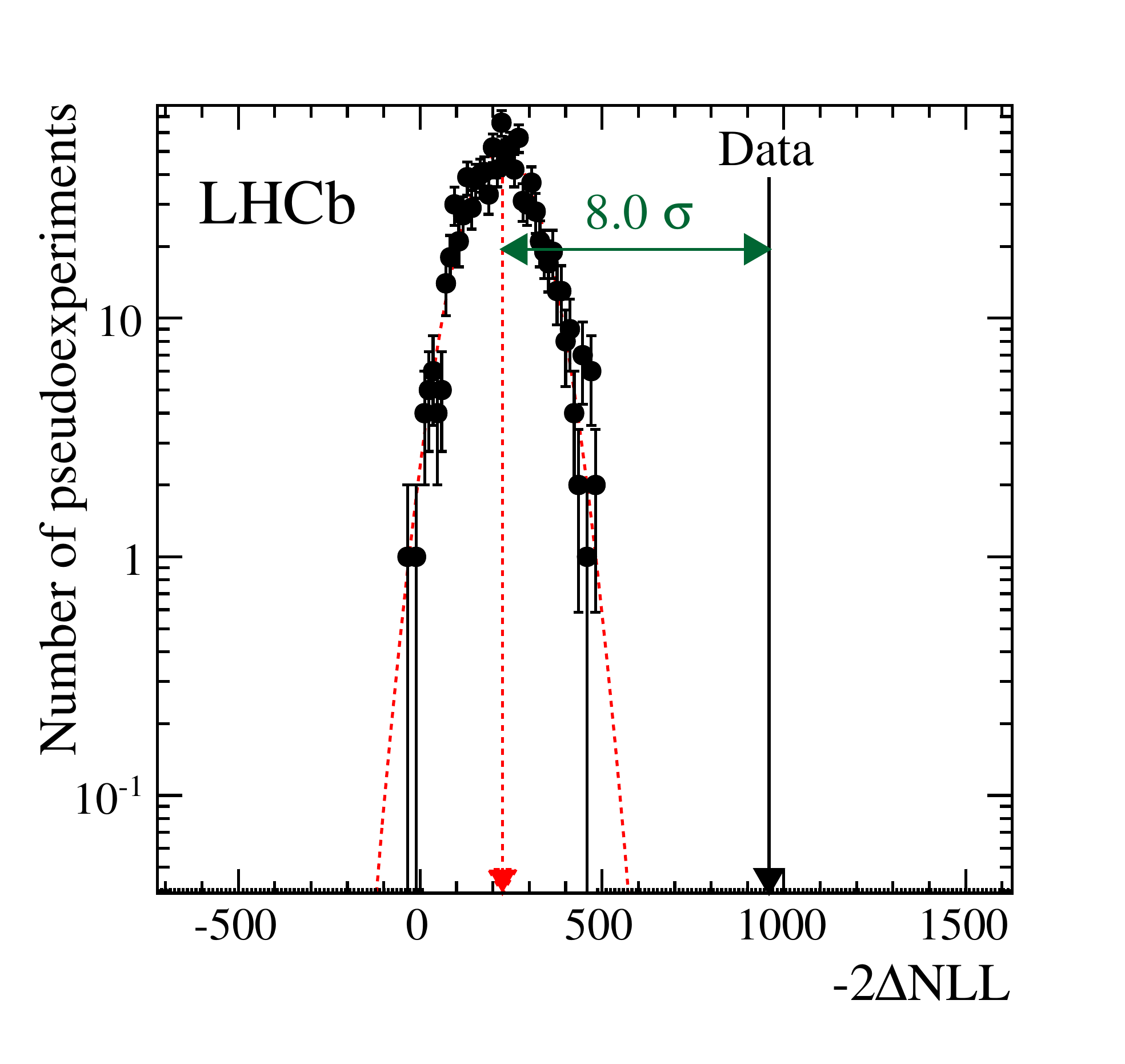}
\end{subfigure}%
\begin{subfigure}[b]{.33\linewidth}
\includegraphics[width=\linewidth]%
{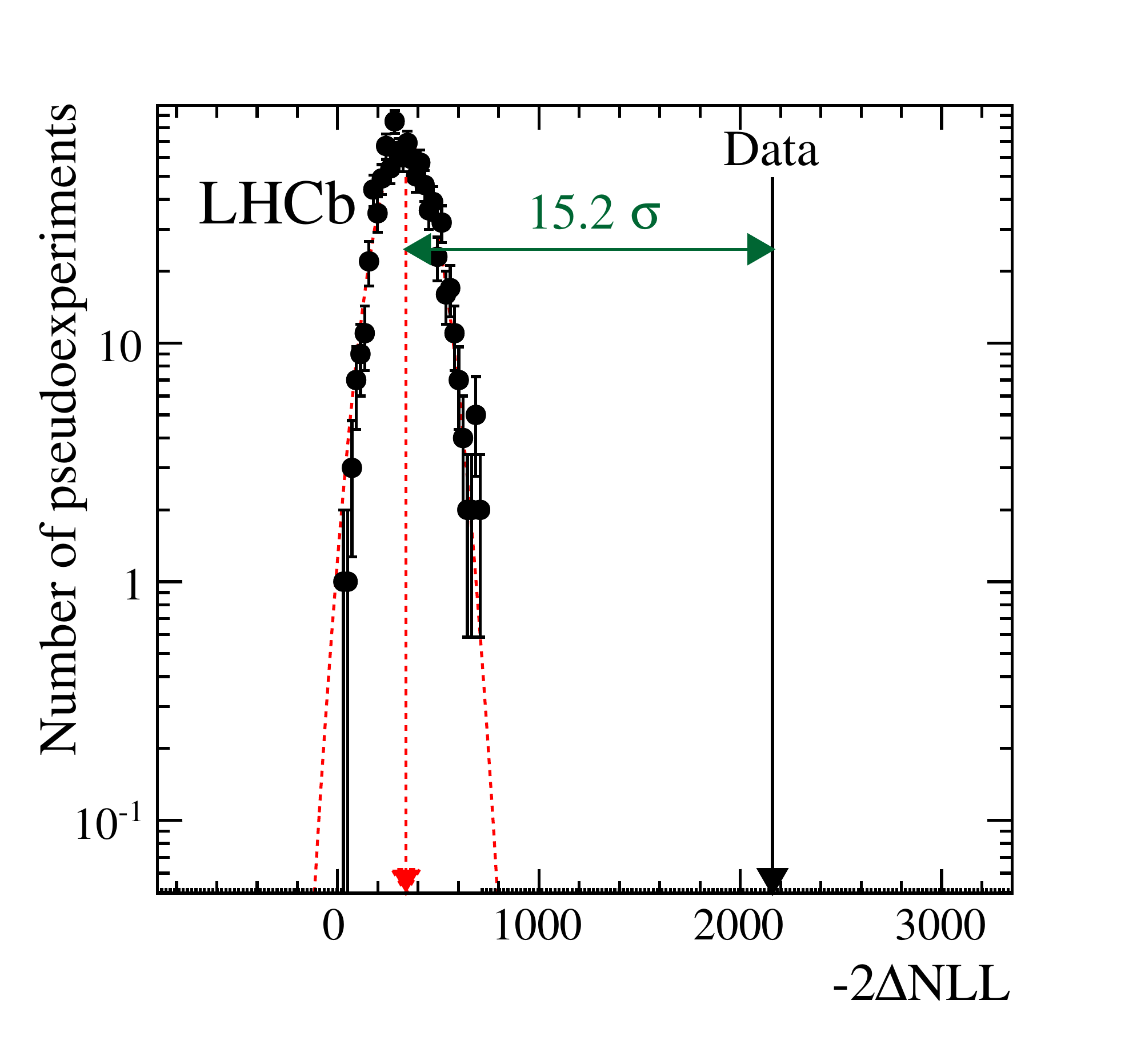}
\end{subfigure}

\caption{ \small Distributions of $-2\NLL$ for the pseudoexperiments (black dots), fitted with a Gaussian function (dashed red line), 
in three different configurations of the \kpi system angular contributions: (left) $\lmax=4$, (middle) $\lmax=6$ and (right) $\lmax$ variable according 
to \autoref{eq:slices}. The black arrow represents the $-2\NLL$ value obtained on data.}
\label{fig:nll-fullspectrum}
\end{figure}

\begin{figure}[t]
\centering
\begin{subfigure}[b]{.33\linewidth}
\includegraphics[width=\linewidth]%
{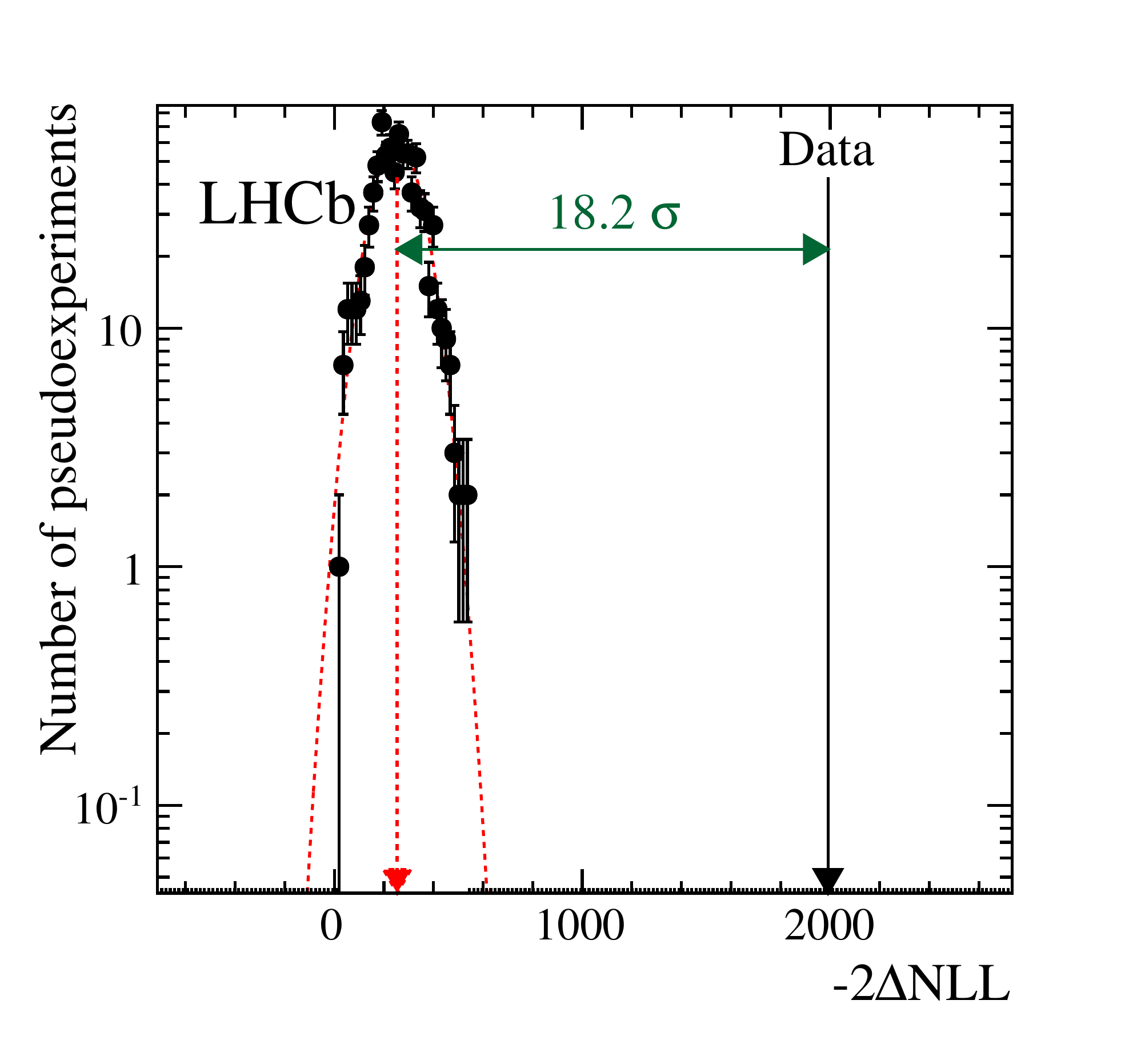}
\end{subfigure}%
\begin{subfigure}[b]{.33\linewidth}
\includegraphics[width=\linewidth]%
{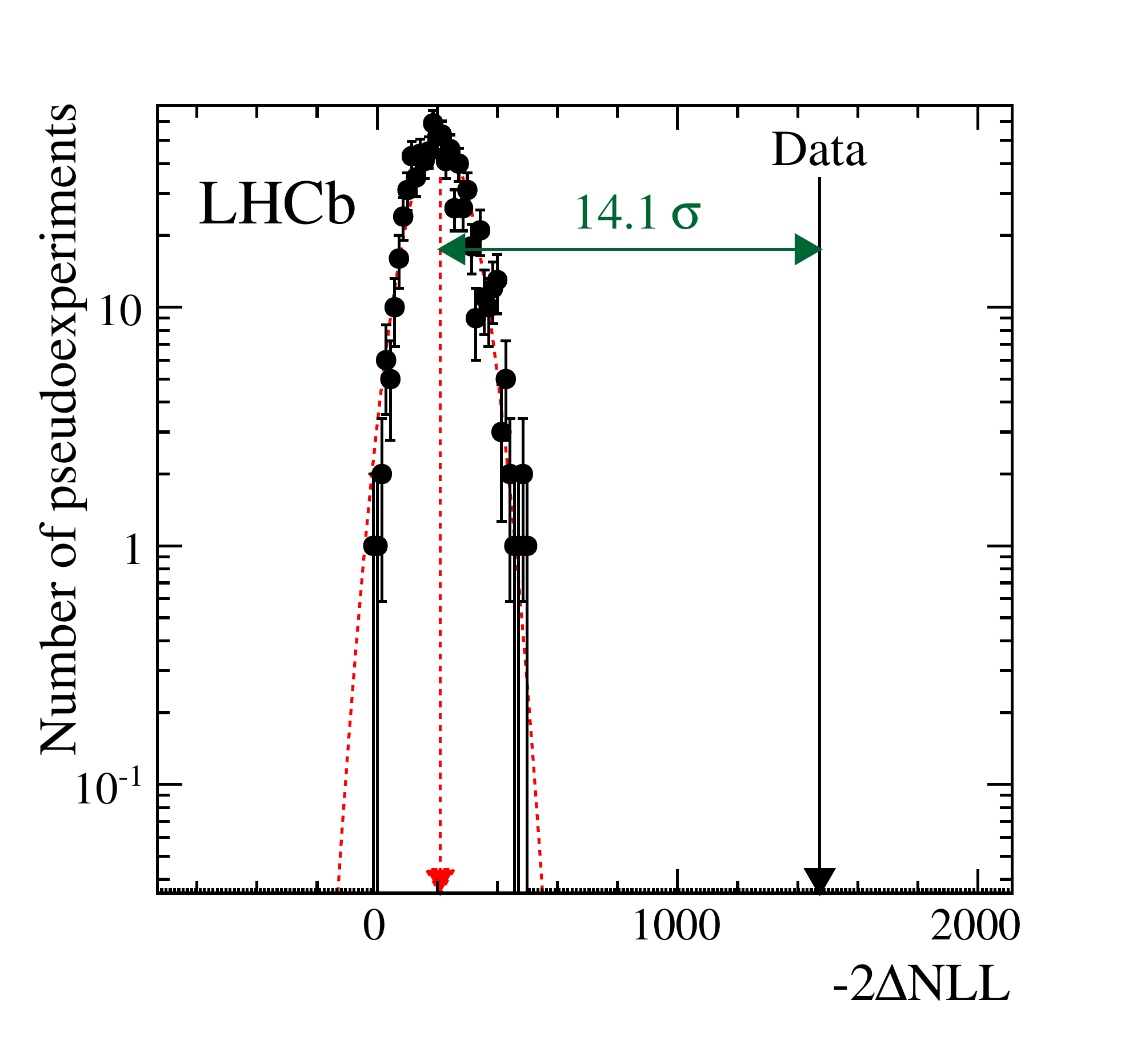}
\end{subfigure}%
\begin{subfigure}[b]{.33\linewidth}
\includegraphics[width=\linewidth]%
{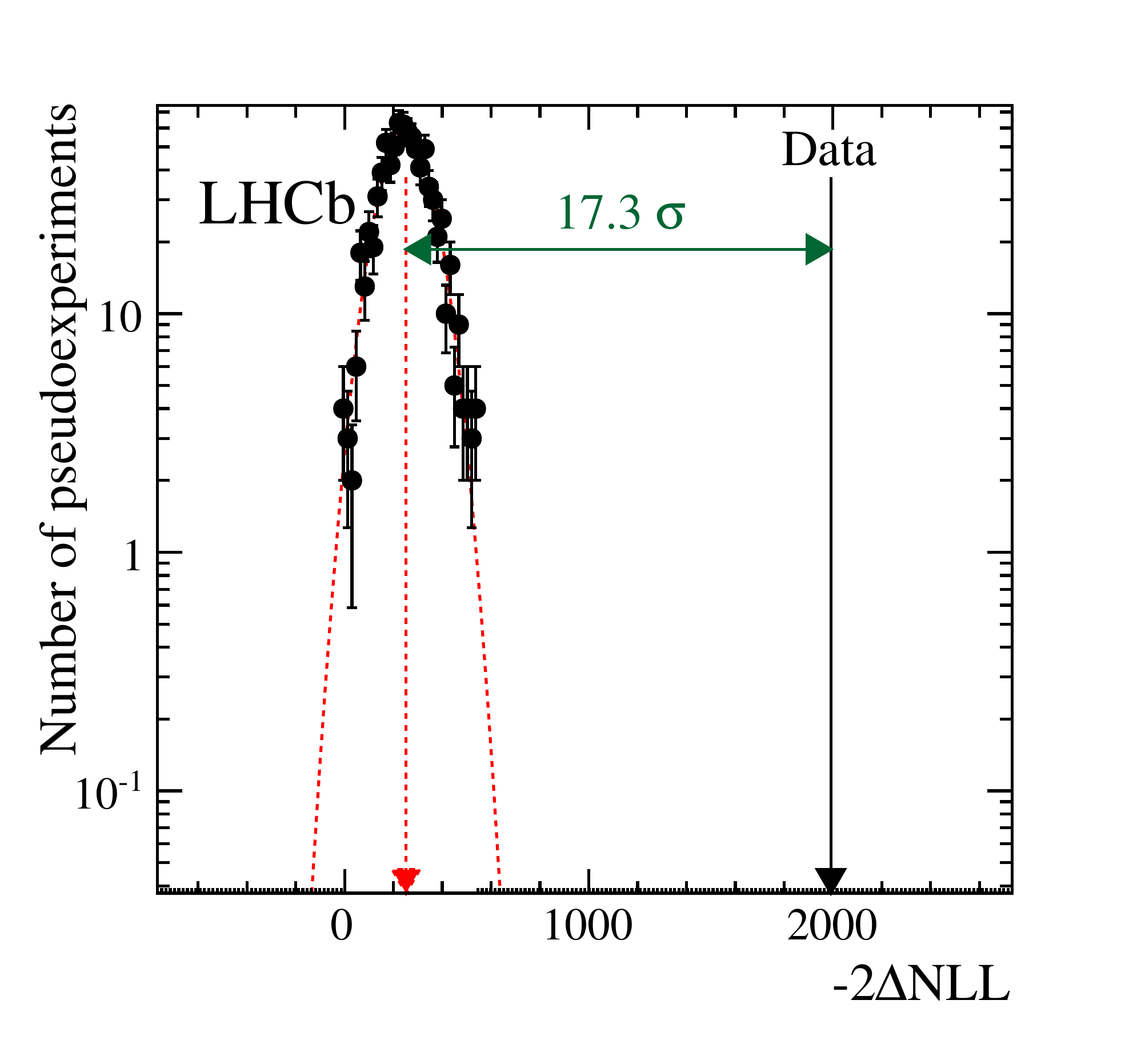}
\end{subfigure}
\caption{ \small Distributions of $-2\NLL$ for the pseudoexperiments (black dots), fitted with a Gaussian function (dashed red line), 
for the region $1000\mevcc<\mkpi<1390\mevcc$ in three different configurations of the \kpi system angular contributions: (left) $\lmax=4$, (middle) $\lmax=6$ and (right) $\lmax$ variable according to \autoref{eq:slices}. The black arrow represents the $-2\NLL$ value obtained on data.}
\label{fig:nll-slice}
\end{figure}

\begin{table}[t]
    \centering
  \begin{tabular}{@{\extracolsep{\fill}}lcc} 
\hline
		 & {\it S}, whole $\mkpi$ spectrum  & {\it S}, $1.0 < \mkpi < 1.39\gevcc$   \\ 
\hline
    $\lmax=4$       &13.3$\sigma$ & 18.2$\sigma$\\
    $\lmax=6$       & \phantom{0}8.0$\sigma$ & 14.1$\sigma$\\
$\lmax(\mkpi)$      &15.2$\sigma$ & 17.3$\sigma$\\
\hline             
    \end{tabular}
    \caption{\small Significance, {\it S}, in units of standard deviations, at which the hypothesis that $\mpsipp$ data can be described
		as a reflection of the $\Kpi$ system angular structure is excluded, for different 
		configurations of the $\Kpi$ system angular contributions. In the second column the whole $\mkpi$ spectrum has been analyzed while in the third one the specified $\mkpi$ cut is applied. }
    \label{tab:significance}
  \end{table}
  
The Monte Carlo method described in \autoref{sec:MIReflectionOnMpsipi} is used for each pseudoexperiment  to produce an $\mpsipp$ probability density function, 
$\mathcal{F}_{\lmax}$, for each of the three $\lmax$ configurations.
To test for the presence of possible contributions from the $\psipp$ dynamics, which are expected to be present in moments of all orders, a fourth configuration is introduced by setting $\lmax$ to the unphysically large value of 30.
By including moments up to $\lmax=30$, most of the features of the $\mpsipp$ spectrum in data are well described, as can be seen in \autoref{fig:lmax30}.
The logarithm of the likelihood ratio is used to define the test statistic

\[ -2{\NLL}_{\lmax} = -2 \sum_{i=1} \frac{W^i_{\text{signal}}}{\Pepsilon^i}\log{\frac{\mathcal{F}_{\lmax}(\mpsipp^i)}{\mathcal{F}_{30}(\mpsipp^i)}} ,\]
\noindent where the sum runs over the events in the pseudo or real experiments.
\par An exotic state in the $\psipp$ system
would give contributions to all \Kpi Legendre polynomial moments, whereas the conventional \Kpi resonances contribute only to moments corresponding to their spin and their interferences. If, for instance, the $\BzToKPiPsi$ decay  proceeds through  S, P and D \Kpi resonances, then
only  moments with  $\lmax\leq4$ would exhibit significant activity. Therefore,  activity in moments of order $\lmax>4$ would suggest the presence of other resonant states contributing to the decay. 
Lower-order \Kpi Legendre polynomial moments, determined from data and used to build the prediction, although strongly dominated by the conventional \Kpi resonances, could also contain a contribution from the exotic state. 
As a consequence, a relatively small $\psipp$ resonant contribution could be accommodated by the Monte Carlo prediction. Conversely, a significant disagreement would imply that the $\psipp$ invariant mass spectrum cannot be explained as a reflection of the activity of known resonances in the \Kpi system, and would therefore constitute strong evidence for the presence of exotic states in the decay $\BzToKPiPsi$.
  
The $\NLL_{\lmax}$ distributions of the pseudoexperiments are shown in \autoref{fig:nll-fullspectrum} (points with error bars)  for each of the three $\lmax$ settings. They are consistent with Gaussian distributions.
The statistical significance, ${\it S}$, to rule out the different hypotheses is the distance, in units of standard deviations, between the mean value of the  $\NLL_{\lmax}$ (dashed red arrow in \autoref{fig:nll-fullspectrum}) and the observed value of the real experiment (continuous black arrow in \autoref{fig:nll-fullspectrum}). 
This ranges from 8 to 15 standard deviations, as listed in \autoref{tab:significance}.

The table also gives the statistical significance obtained by  
restricting the analysis to the region $1000\mevcc<\mkpi<1390\mevcc$, where 
the presence of the structure around the \Zp mass is most evident, as shown in \autoref{fig:nll-slice}. 
Thus, the hypothesis that the data can be explained solely in terms of plausible
\Kpi degrees of freedom can be ruled out without making any assumption on the exact shapes of the \Kpi resonances present and their interference patterns.



\ifthenelse{\boolean{prl}}{ }{ \tempclearpage }

\ifthenelse{\boolean{prl}}{ }{ \tempclearpage }
\section{Summary and conclusions}
\label{sec:summary}

A satisfactory description of the $\psipp$ mass spectrum in the decay $\BzToKPiPsi$ cannot be obtained solely from the reflections of the angular structure of the \Kpi system.
In particular, a clear peaking structure in the $4430\mevcc$ mass region remains unexplained.  
Through a hypothesis-testing procedure based on the likelihood-ratio estimator, compatibility between the data and predictions taking
into account the reflections of \Kpi states up to spin three, is excluded with a significance exceeding 8$\sigma$.
The most plausible configuration, which allows $\Kpi$ states with spin values depending on the \Kpi mass, is excluded with a significance of more than 15$\sigma$.
\par This work represents an alternative and model-independent confirmation of the existence of a $\psipp$ resonance in the same mass region in which previous model-dependent amplitude analyses have found signals
\cite{Choi:2007wga,Chilikin:2013tch,LHCb-PAPER-2014-014}. 



\ifthenelse{\boolean{prl}}{ }{ \tempclearpage }
\section*{Acknowledgements}
 
\noindent We express our gratitude to our colleagues in the CERN
accelerator departments for the excellent performance of the LHC. We
thank the technical and administrative staff at the LHCb
institutes. We acknowledge support from CERN and from the national
agencies: CAPES, CNPq, FAPERJ and FINEP (Brazil); NSFC (China);
CNRS/IN2P3 (France); BMBF, DFG and MPG (Germany); INFN (Italy); 
FOM and NWO (The Netherlands); MNiSW and NCN (Poland); MEN/IFA (Romania); 
MinES and FANO (Russia); MinECo (Spain); SNSF and SER (Switzerland); 
NASU (Ukraine); STFC (United Kingdom); NSF (USA).
We acknowledge the computing resources that are provided by CERN, IN2P3 (France), KIT and DESY (Germany), INFN (Italy), SURF (The Netherlands), PIC (Spain), GridPP (United Kingdom), RRCKI (Russia), CSCS (Switzerland), IFIN-HH (Romania), CBPF (Brazil), PL-GRID (Poland) and OSC (USA). We are indebted to the communities behind the multiple open 
source software packages on which we depend. We are also thankful for the 
computing resources and the access to software R\&D tools provided by Yandex LLC (Russia).
Individual groups or members have received support from AvH Foundation (Germany),
EPLANET, Marie Sk\l{}odowska-Curie Actions and ERC (European Union), 
Conseil G\'{e}n\'{e}ral de Haute-Savoie, Labex ENIGMASS and OCEVU, 
R\'{e}gion Auvergne (France), RFBR (Russia), XuntaGal and GENCAT (Spain), The Royal Society 
and Royal Commission for the Exhibition of 1851 (United Kingdom).


\ifthenelse{\boolean{prl}}{ }{ \clearpage }

\addcontentsline{toc}{section}{References}
\ifx\mcitethebibliography\mciteundefinedmacro
\PackageError{LHCb.bst}{mciteplus.sty has not been loaded}
{This bibstyle requires the use of the mciteplus package.}\fi
\providecommand{\href}[2]{#2}

\newpage
\centerline{\large\bf LHCb collaboration}
\begin{flushleft}
\small
R.~Aaij$^{38}$, 
B.~Adeva$^{37}$, 
M.~Adinolfi$^{46}$, 
A.~Affolder$^{52}$, 
Z.~Ajaltouni$^{5}$, 
S.~Akar$^{6}$, 
J.~Albrecht$^{9}$, 
F.~Alessio$^{38}$, 
M.~Alexander$^{51}$, 
S.~Ali$^{41}$
% ...

{\footnotesize \it
% ...
$ ^{5}$Clermont Universit\'{e}, Universit\'{e} Blaise Pascal, CNRS/IN2P3, LPC, Clermont-Ferrand, France\\
$ ^{6}$CPPM, Aix-Marseille Universit\'{e}, CNRS/IN2P3, Marseille, France\\
% ...
$ ^{9}$Fakult\"{a}t Physik, Technische Universit\"{a}t Dortmund, Dortmund, Germany\\
% ...
$ ^{37}$Universidad de Santiago de Compostela, Santiago de Compostela, Spain\\
$ ^{38}$European Organization for Nuclear Research (CERN), Geneva, Switzerland\\
% ...
$ ^{41}$Nikhef National Institute for Subatomic Physics, Amsterdam, The Netherlands\\
% ...
$ ^{46}$H.H. Wills Physics Laboratory, University of Bristol, Bristol, United Kingdom\\
% ...
$ ^{51}$School of Physics and Astronomy, University of Glasgow, Glasgow, United Kingdom\\
$ ^{52}$Oliver Lodge Laboratory, University of Liverpool, Liverpool, United Kingdom\\
% ...
}
\end{flushleft}


\newpage



\end{document}